\documentclass[acmsmall]{acmart}
\usepackage{url}
\usepackage{hyperref}
\AtBeginDocument{%
  \providecommand\BibTeX{{%
    \normalfont B\kern-0.5em{\scshape i\kern-0.25em b}\kern-0.8em\TeX}}}

\setcopyright{acmcopyright}
\copyrightyear{2018}
\acmYear{2023}
\acmDOI{10.1145/1122445.1122456}

\acmJournal{CSUR}
\acmVolume{00}
\acmNumber{0}
\acmArticle{000}
\acmMonth{2}

\begin{document}

\title{CAPTCHA Types and Breaking Techniques: Design Issues, Challenges, and Future Research Directions}

\author{Noshina Tariq}
\email{noshina.tariq@mail.au.edu.pk}
\orcid{1234-5678-9012}
\affiliation{Department of Avionics Engineering,
  \institution{ Air University}
  \streetaddress{Service Road E-9/E-8}
  \city{Islamabad}
  \country{Pakistan}}
  
\author{Farrukh Aslam Khan}
\authornotemark[1]
\affiliation{Center of Excellence in Information Assurance,
  \institution{King Saud University}
  \streetaddress{P.O. Box  92144}
  \city{Riyadh, 11653}  
  \country{Saudi Arabia}}
\email{fakhan@ksu.edu.sa}

\author{Syed Atif Moqurrab}
\email{atif@gachon.ac.kr}
\affiliation{%
  \institution{School of Computing, Gachon University}
  \streetaddress{1342, Seongnam-daero, Sujeong-gu, Seongnam-si, 13120}
  \city{Seongnam}
  \country{South Korea}}
  \author{Gautam Srivastava}
\email{srivastavag@brandonu.ca}
\affiliation{%
  \institution{Brandon University}
  \streetaddress{Brandon, MB R7A 6A9,}
  \city{Brandon}
  \country{Canada}}

\renewcommand{\shortauthors}{Noshina et al.}

\begin{abstract}
 The proliferation of the Internet and mobile devices has resulted in malicious bots’ access to genuine resources and data. Bots may instigate phishing, unauthorized access, denial-of-service, and spoofing attacks, to mention a few. Authentication and testing mechanisms to verify the end-users and prohibit malicious programs from infiltrating the services and data are strong defense systems against malicious bots. Completely Automated Public Turing test to tell Computers and Humans Apart (CAPTCHA) is an authentication process to confirm that the user is a human; hence, access is granted. This paper provides an in-depth survey on CAPTCHAs and focuses on two main things: (1) a detailed discussion on various CAPTCHA types along with their advantages, disadvantages, and design recommendations, and (2) an in-depth analysis of different CAPTCHA breaking techniques. The survey is based on over two hundred studies on the subject matter conducted since 2003 to date. The analysis reinforces the need to design more attack-resistant CAPTCHAs  while keeping their usability intact. The paper also highlights the design challenges and open issues related to CAPTCHAs. Furthermore, it also provides useful recommendations for breaking CAPTCHAs.   
\end{abstract}



\maketitle
\section{Introduction}
The Internet is a primary mode of communication in the modern age of the Internet of Things (IoT)~\cite{tariq2021blockchain}. The availability of a diverse range of mobile devices and more affordable high-speed data subscriptions boosted consumers' interest in Internet use. From its inception, Internet security has been the primary concern of web developers. As the Internet continues to grow for providing various websites, services, and blogs likewise grow. Today's websites are dedicated to the public, transportation, entertainment, financial services, food items, healthcare, and hotel reservations, to mention a few. In this regard, the growing user base also necessitates the deployment of high-end computing power at the websites~\cite{derhab2021tweet}. However, these high-end processors are rendered ineffective if the high-end automated device targets them. Therefore, the defense against such automated attacks is imperative. However in an open-access network environment, the ubiquity of the Internet is encouraging security threats for individuals gaining network access. It is not an easy task for the web service providers to understand whether a certain application is accessed by a bot or a human user~\cite{tanthavech2019captcha}. 

Nonetheless, bot programs are extremely useful to carry out cyclic and time intense operations. Oppressed for malevolent work, bot programs have human behavior reproduction ability. Therefore, security is the main ingredient in a reliable communication interface~\cite{chow2019CAPTCHA}. It points to sheltering and protecting networks from vicious attacks. Introducing security services in computer networks seeks to deliver convenient, reliable, and classified application environment~\cite{ul2021ctrust}~\cite{ahmed2018malicious}~\cite{imran2019toward}, and~\cite{ouyang2021cloud}. There is various computer exploitation, including worms, spoofing, viruses, and network crossing. Moreover, other kinds of attacks would be unauthorized ones: covert channels, Denial of Services (DoS), and compromised and malicious nodes, to name a few~\cite{gonen2018secure}. A majority of Completely Automated Public Turning test to tell Computers and Humans Apart (CAPTCHA) systems have been proposed that use various features, such as characters, images, and audio, to generate challenges that effectively prevent automated bots. However, recent advancements in Artificial Intelligence (AI) in general and Computer Vision (CV) in particular have greatly improved automated systems' ability to solve such tasks~\cite{wang2023experimental,zhang2023secure}. As a consequence, almost all conventional CAPTCHA systems have been compromised. Hence, there needs to have an in-depth analysis of CAPTCHA formation and its breaking techniques to make them more robust and usable.


CAPTCHA (AKA Human Interaction Proof (HIP)) is introduced to mitigate spurious network resources access~\cite{wang2023extended,tariq2019securing,kumar2021benchmarks}. It was coined in 2000 by Luis Von Ahn at Carnegie Mellon University~\cite{tariq2018match}. It is a protection method for avoiding electronic registration, spam, and harmful bot applications. They are well-known for providing practical security by differentiating humans from computers; for example, lock-free Email services, shielding online polling against online voting bots, or malicious email sign-ups. They are effective in handling email worms and spam and in preventing dictionary-based password attacks~\cite{von2003CAPTCHA}. In general, a test that is simple for people to solve and difficult for computers is generated and evaluated. It is deemed successful if its success rate in human solutions is greater than 90\%, and computers reach just a success rate of less than 1\%~\cite{lazar2012soundsright}. A strong CAPTCHA is generally known, not only functional but stable. It is continuously evolving technology and is increasingly growing in literature and practice. The emerging literature references must be summarised, and a more comprehensive analysis of CAPTCHAs must be carried out. 

\begin{figure}[t]
\centering
\includegraphics[width=3in]{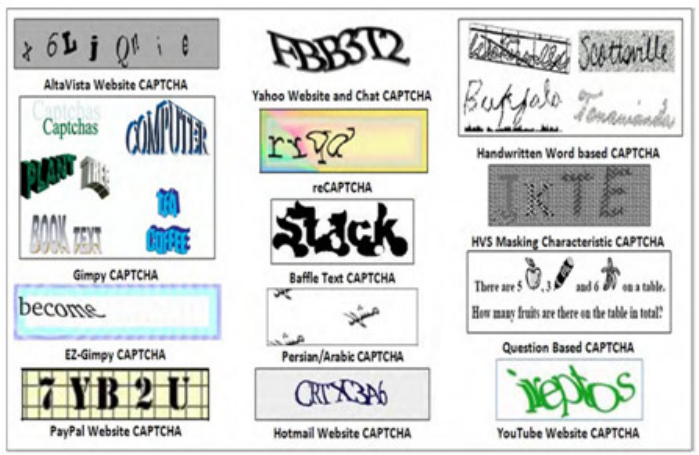}
\caption{Different kinds of text CAPTCHAs.}\label{1}
\end{figure}

Various models have appeared after CAPTCHAs have been produced. Among different types of CAPTCHAs, text-based CAPTCHAs are widely used~\cite{chow2019CAPTCHA}, Fig.~\ref{1} shows some examples of text-based CAPTCHAs~\cite{banday2009image}. However, researchers attack CAPTCHAs to identify which interface features are good for its protection to make it resilient. This paper presents an in-depth and exhaustive study of different CAPTCHA schemes and their breaking techniques proposed from 2003 to date. We analyzed its types regarding their strengths and weaknesses. This paper also presents the major applications of CAPTCHAs. We also studied several CAPTCHA breaking techniques, which are discussed later in the paper. Based on our comprehensive study and analysis, we classified CAPTCHA types into OCR and non-OCR categories presented in Section~\ref{tc}. We also classified CAPTCHA breaking techniques into four categories discussed in Section~\ref{cb}. Later, we present several challenges and open issues related to CAPTCHA designs and breaking.
 
  This paper has the following major contributions:
   \begin{enumerate}
   \item The classifications (taxonomies) of CAPTCHA types and breaking techniques is presented and thoroughly analyzed.
   \item An analysis of state-of-the-art CAPTCHA survey papers (for both construction and breaking) is done, highlighting their contributions and missing research areas in the said domain.
   \item CAPTCHA practices and applications are discussed, helping readers better understanding the areas where CAPTCHAs may be useful.
       \item Based on CAPTCHA formation, exhaustive state-of-the-art study and analysis are presented. In line with CAPTCHA types, advantages and disadvantages accordant with the proposed taxonomies are also provided for engineering better designs.
       \item Based on CAPTCHA breaking techniques, a comprehensive state-of-the-art study and analysis are presented to fix the loopholes in CAPTCHA designs that may be exploited by the attackers.
       \item In-depth analysis and future prospects of design issues and challenges are imparted encompassing a prolific discussion on CAPTCHA sturdiness and usability.
   \end{enumerate}

   In this survey paper, Section~\ref{rw} presents related work that provides a comparison among state-of-the-art survey. In Section~\ref{cp}, important CAPTCHA application areas are discussed. Section~\ref{tc} presents different CAPTCHA techniques based on their formation and design. Challenges and open issues in CAPTCHAs designs are discussed in Section~\ref{coi}. Section~\ref{cb} represents the survey of state-of-the-art CAPTCHA breaking algorithms. Section~\ref{coi1} presents a productive discussion on CAPTCHA robustness and usability along with the challenges and open issues, and Section~\ref{con} gives the conclusion and future directions.

\section{Related surveys}\label{rw}
This section briefly discusses state-of-the-art surveys from 2010 to date. Table~\ref{zzz} presents a comparison between state-of-the-art and this paper. We compared contributions and the subject matter, such as the survey about CAPTCHA types, attacks on CAPTCHAs or both. It is also investigated that the survey has proposed any taxonomy of CAPTCHA types or its breaking techniques or not. That helps readers to understand the subject matter in a modular way. It is also essential that a survey discusses challenges and open issues associated with the subject matter from our perspective. That helps the readers to find the gaps and future research directions.

\begin{table*}
\centering
\tiny
\caption{Comparison among state-of-the-art surveys}
\label{zzz}       
\begin{tabular}{|c| c| c| c| c| c| c| c|}
\hline
Ref. & Year& Type-based& Attack-based & Applications&Taxonomy& Challenges \& Open issues & Recommendations \\ \hline

This paper&2023&\checkmark&\checkmark & \checkmark&\checkmark& \checkmark & \checkmark \\ \hline
\cite{alsuhibany2023survey}&2023&\checkmark&$\times$&$\times$& $\times$& $\times$ & $\times$ \\ \hline
\cite{kumar2022systematic}&2022&\checkmark&\checkmark&\checkmark& $\times$& $\times$ & $\times$ \\ \hline
\cite{magdy2021comprehensive}&2021&\checkmark&\checkmark&$\times$& \checkmark& $\times$ & $\times$ \\ \hline
\cite{kumar2021systematic}&2021&\checkmark&\checkmark& \checkmark&$\times$& $\times$ & $\times$\\ \hline
 \cite{srivastava2021survey} &2021&$\times$&\checkmark &$\times$& $\times$& $\times$ &$\times$\\ \hline
 \cite{kumar2021captcha} &2021&\checkmark&$\times$ & $\checkmark$&$\times$& $\times$& $\times$\\ \hline
 \cite{Zhang2019survey} &2019&\checkmark&\checkmark & $\times$&$\times$& \checkmark &$\times$\\ \hline
 \cite{xu2020survey}&2020&\checkmark&\checkmark&$\times$&$\times$ &$\times$ &$\times$ \\ \hline
 \cite{dionysiou2020sok}&2020&$\times$ &\checkmark&$\times$&$\times$ &$\times$ &$\times$\\ \hline
 \cite{divyashree2016survey} & 2016& \checkmark& $\times$ &\checkmark&$\times$& $\times$ &$\times$ \\ \hline
 \cite{guerar2018completely} & 2018& \checkmark& $\times$ &$\times$&$\times$&$\times$ &$\times$\\ \hline
 \cite{greene2018large} & 2018& \checkmark& \checkmark &$\times$&$\times$& $\times$& $\times$ \\ \hline
  \cite{roshanbin2013survey} & 2013& \checkmark& \checkmark &$\times$&$\times$&\checkmark &$\times$\\ \hline
  \cite{moradi2015captcha} & 2015& \checkmark& $\times$ &$\times$&$\times$&$\times$ & $\times$ \\ \hline
   \cite{bursztein2010good} & 2010& \checkmark& $\times$ &$\times$&$\times$& $\times$& $\times$ \\ \hline
 \cite{banday2011study} & 2011& \checkmark& \checkmark &$\times$&$\times$&$\times$ &$\times$\\ 
\hline
\end{tabular}
\end{table*}
Magdy et al.~\cite{magdy2021comprehensive} presented a survey on CAPTCHA types and breaking techniques. Though they have discussed its different types and limited usability issues; however, the survey is not comprehensive. Kumar et al.~\cite{kumar2021systematic} presented a systematic survey on CAPTCHAs. This paper encompasses a good discussion of different CAPTCHAs and their breaking. However, types and attacks are not classified as the techniques used in their breaking. Moreover, it does not discuss challenges and open issues in CAPTCHA designing and breaking. However, our paper provides a more comprehensive survey on CAPTCHA types and breaking techniques. Still, it also discusses challenges and open issues in this regard, along with the recommendations for designing and breaking CAPTCHAs. Zhang et al.~\cite{Zhang2019survey} discuss the advancements made scientifically and technologically on CAPTCHAs' design and attacks. They discussed the usability, robustness, and limitations of different types of CAPTCHAs. Besides, the attack techniques and differences between the formation of CAPTCHAs are categorically debated. This paper discusses recommendations for future studies and suggests several issues that require extra analysis. However, a taxonomy of types and breaking techniques is not presented. 

The CAPTCHA usability and its progression technology are also assessed in~\cite{xu2020survey}. It examines the CAPTCHA mechanisms from usability and security aspects, unlike other studies that have been done on it. The introduction of new CAPTCHA formats (game CAPTCHA) and attacking methods (deep learning-based attacking) are also included in this paper. However, they do not present a taxonomy and a discussion about challenges and open issues. Besides,~\cite {dionysiou2020sok} carries out methodical investigation and categorization. The study is based on the ultramodern ML-based method for self-operating text-based CAPTCHA breaking issues. An investigation into the present status and robustness of text-based CAPTCHA is also presented. The present-day Internet users use it to counter ML-based automated breaking tools in this paper. This literature also recommends that ML-based as a breaking tool is very efficacious because of its speed, precision, and reverie in the CAPTCHA solution.  This paper encourages combatting the issues associated with current internet services and ML-based methods; a reverse Turing test is encouraged in this paper. Reverse Turing systems are recommended because they offer less stress to human users and give computer/robots challenging problems. However, the paper does not present a taxonomy of the CAPTCHA types or their breaking techniques. It also lacks in discussing challenges and open issues in this regard.


CAPTCHA types, their usability, and deficiencies are discussed in ~\cite{divyashree2016survey}. However, this survey focuses on CAPTCHA types only. Moreover, it does not discuss attacks made on CAPTCHAs. It also does not present a taxonomy and the challenges and open issues associated with CAPTCHA types. \cite{guerar2018completely} compares CAPTCHA and completely automated public physical test to tell computers and humans apart (CAPPCHA). CAPPCHA and CAPTCHA are alike because they are both used to protect websites from abusive automated programs. As discussed in this paper, CAPPCHA improves the authentication technique required for mobile devices. Besides, a proof of concept is also provided. However, an extensive analysis of the subject must be carried out to confirm the efficiency of CAPPCHA. Therefore, this paper aims to thoroughly investigate CAPTCHA's officiousness and its usability involving several human users. However, this paper also does not talk about CAPTCHA breaking techniques, taxonomy, and challenges.

In~\cite{greene2018large}, keywords, such as; login, cart, registration, account, post, auth, join, register, sign, password, and subscribe, were used by a personally web designed crawler. The crawler makes use of a beautiful soup library to investigate 30,000 websites and check if they use CAPTCHA to protect their URLs. This paper also examines and categorizes CAPTCHA as regards its limitations. Out of 30,000 web pages, only 10,017 uses CAPTCHA. According to the facts provided by this survey, the most used CAPTCHA includes; text/image-based CAPTCHA, reCAPTCHA, text-based CAPTCHA, math, slider, FunCAPTCHA, custom, and audio-based CAPTCHA. However, taxonomy and open issues are not addressed in this paper.

\cite{roshanbin2013survey} discusses three essential things. They include; recent CAPTCHAs and associative attacks, the inquiry into the vigor and usefulness of recently introduced CAPTCHAs, and propose new techniques to improve recently introduced CAPTCHAs. However, the paper does not suggest a taxonomy. \cite{moradi2015captcha}~explores substitutes of CAPTCHA, usability, and services provide researchers with enough information so that new techniques and solutions can be suggested. Thus far, the positive strides made in CAPTCHA advancements, various CAPTCHAs substitutes, were categorized, evaluated, and compared with their shortcomings. Solutions to the shortcoming were also proposed. However, CAPTCHA breaking is out of the scope of this paper. Besides, there is no proposed taxonomy, and no discussion on open issues are discussed either.

From the human standpoint,~\cite{bursztein2010good} extensively reviews CAPTCHAs to gauge how much an average user is affected by CAPTCHA. However, this paper focuses only on CAPTCHA types. There is no taxonomy presented. There is no discussion on open issues and challenges presented in the paper. Similarly, Bandy et al.~\cite{banday2011study} presented a classification of CAPTCHAs and compared its different kinds. The CAPTCHAs are classified and compared as regards their usability and security. Conventional methods for generating and breaking text and image-based CAPTCHA are also introduced. Lastly, this paper recommends various ways to enhance vigor, usability and to tackle security issues. However, taxonomy and open issues are not discussed.

\section{CAPTCHA Practices} \label{cp}
There are several CAPTCHA practices and applications; this section highlights some of the most common practised CAPTCHA applications below: 

\subsection{Mitigating e-mail (address) security and spams}
Ahn et al.~\cite{von2004labeling} described how web scrapers (programs that extract data from websites) might be prevented from obtaining users' e-mail addresses by requiring them to answer a CAPTCHA prior to revealing their e-mail addresses. It may also be used for the prohibition on the use of fake e-mail addresses. Most businesses offer complimentary e-mail services in today's information era. They are the targeted victims of bot assaults. Numerous businesses adopt such methods to obtain complimentary e-mail addresses to distribute spam. A most effective yet simplest way of avoiding unwanted bot assaults is to utilise CAPTCHA-based security. Pope and Kaur~\cite{pope2005human} highlighted that the majority of e-mail providers had implemented such an approach.

\subsection{Phishing Attacks Prevention}
Phishing is a technique used by unethical attackers (e.g., hackers and crackers) to gain access to online banking, e-voting, social media, and e-mail accounts by displaying a similar false web page to the account owner~\cite{kheshaifaty2020preventing}~\cite{tariq2018match}. For instance, these attackers send a link to certain targeted account holders through the e-mail accounts, and by clicking the link, they submit sensitive details, such as usernames and passwords. It results in the loss of confidential information such as official records or project bids. Phishing attackers collect bank information such as usernames, passwords, and credit/debit card information. Certain websites are legitimate financial websites, and some are duped into providing personal and private web to these false websites. These attackers do not utilise their method to break internet accounts, which makes the tracking very challenging. Additionally, cyber law is not sufficient to prosecute such unethical hackers~\cite{kumar2021systematic}. Under such cases, certain countermeasures must be adopted in conjunction with a strong authentication process. Hence, CAPTCHA is inevitable in both bot and phishing attack scenarios.

\subsection{Banning robots from playing games}
Using the CAPTCHA is easier than previously thought to restrict online bot gaming. It is extremely simple to keep robots and computers from playing online games, as found by~\cite{kheshaifaty2020preventing},~\cite{li2018captcha}, and~\cite{golle2005preventing}. 

\subsection{Ensuring a strong internet protection} 
A recent online paper outlined the theory of dictionary attack, which aims to find the passphrase or password of an authentication system by accessing probable alternatives. CAPTCHA is particularly efficient in protecting against such dictionary assaults in these types of circumstances, according to~\cite{chakrabarti2007password},~\cite{alajmi2020password}, and~\cite{khan2020captcha}.

\subsection{A way to handle worms and spam} 
CAPTCHA is used to combat worms and spam and accept mail only if no computer program is present, proving that humans are behind it~\cite{magdy2021comprehensive},~\cite{saxena2017handling}, and~\cite{vaithyasubramanian2019enhancing}.

\subsection{Stronger password security} 
Attempting to enter the password after too many incorrect tries may result in an account being locked, but this is not a superior approach. Ahn et al.~\cite{von2004telling} claim that if a computer program tries to complete a task, it may be overridden by requiring the user to input a CAPTCHA to demonstrate that it is a person and not a computer program.

\subsection{A search engine bot's barrier} 
Pope and Kaur~\cite{pope2005human} and Kim and Luke~\cite{kim2019search} pointed out that if a business wants its web pages to remain undisturbed, CAPTCHA will help to block computers from indexing the sites.

\subsection{Keep automatic online polls to a minimum}
CAPTCHA may also be used to prevent a computer program from voting in online polls. The claims of~\cite{von2004telling} and~\cite{pope2005human} asserted that even while it cannot be used to prevent a person from voting more than once, it cannot be used to allow them to vote more than once either.

\begin{figure}[h]
\centering
\includegraphics[width=4in]{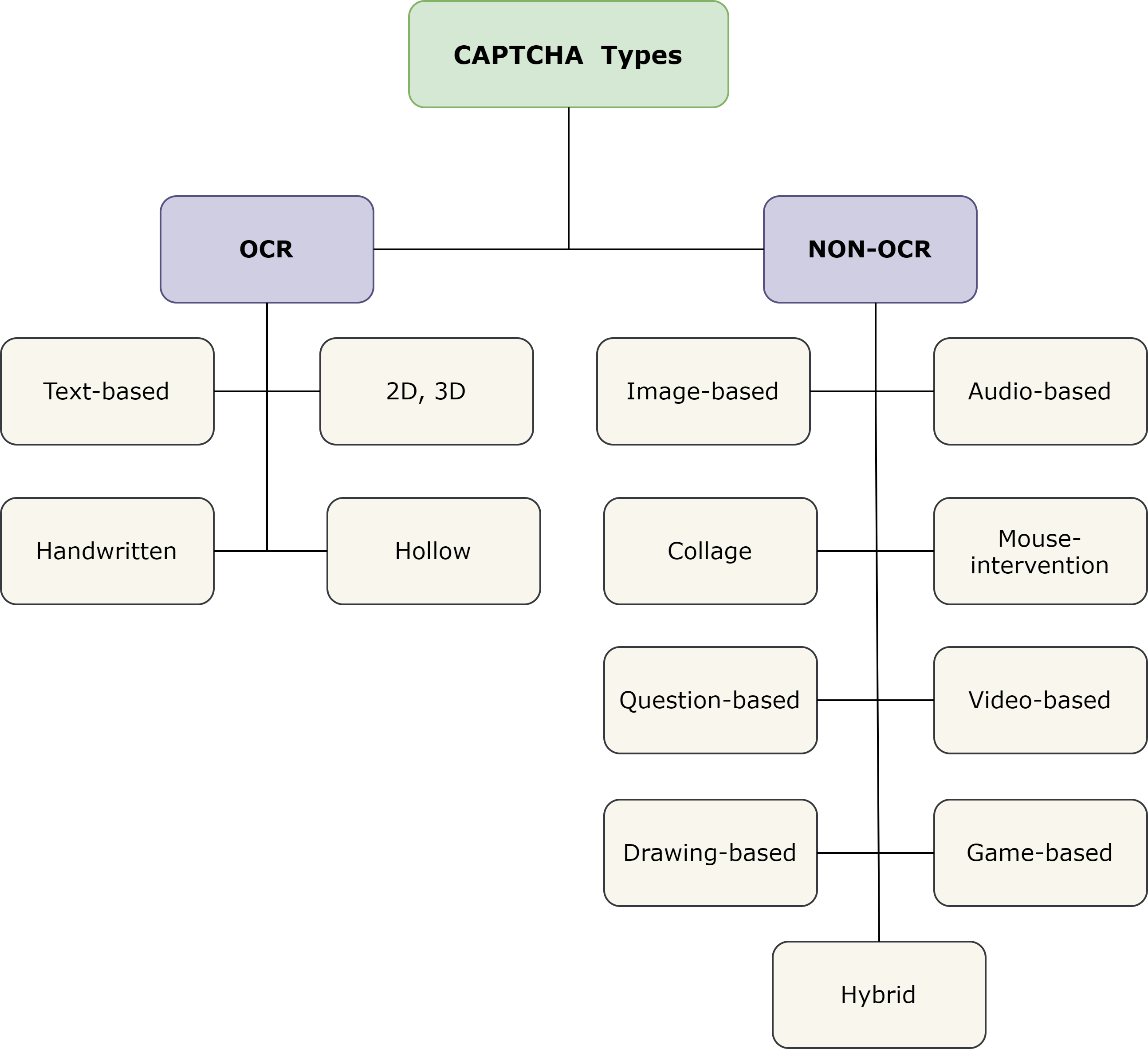}   
\caption{The proposed taxonomy of CAPTCHA types.}\label{2}
\end{figure}

\subsection{A Rejecting Shopping Agent Solution} 
This kind of software is quite popular today: You may build one that provides you with a full cost comparison from many comparable websites, for example, trivago or makemytrip~\cite{kumar2021systematic}. Since the customer cannot view all the ads from these online shops, the online stores lose out. Rejecting this kind of program from disclosing pricing information is extremely successful using the CAPTCHA~\cite{kute2018video}.

\section{Classification of CAPTCHAs based on their types}\label{tc}
     The CAPTCHA schemes may be broadly classified into Optical Character Reader (OCR) and non-OCR CAPTCHA formations. They can further be classified into different types, as shown in Fig.~\ref{2}. This section presents different CAPTCHA schemes classified based on OCR-based and Non-OCR CAPTCHAs, along with each approach's pros and cons.

\subsection{OCR-Based CAPTCHA Schemes}
This section presents the most commonly OCR-based  CAPTCHA schemes.

\subsubsection{Text-based CAPTCHAs}
In 1997, the first bot attack was made on Alta Vista that submitted automated URLs. Therefore, distorted arbitrary text CAPTCHAs were generated by Andrei Border to deter bots~\cite{bandaystudy}.


\begin{figure}[h]
\centering
\includegraphics[width=2in]{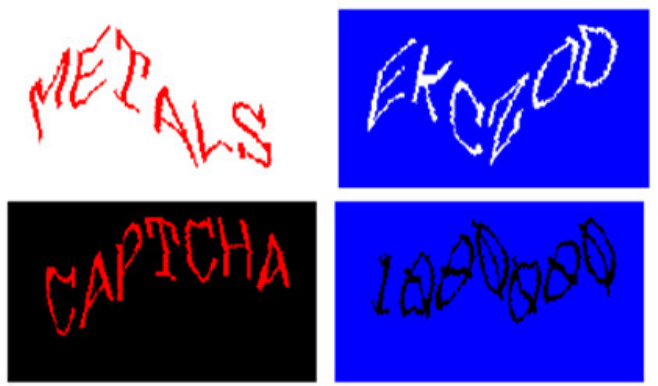}
  \caption{Text CAPTCHAs at CAPTCHAservice.org.}\label{3}
\end{figure}

Text-based CAPTCHAs are categorized as reading-based CAPTCHAs. The user reads a text made up using only letters or a random combination of numbers or letters and inputs them into a given box in the same sequence~\cite{kumar2022usability}. It is quite troublesome for a computerized program to read twisted letters or numbers from clutter. This clutter could consist of anything like a grid, noise, shapes, lines, hues, shades, or contrasting colors. These CAPTCHAs are broadly used by many websites such as Google, rapid-share, MSN, Yahoo, YouTube, and Badongo. An example of a text-based CAPTCHA is shown in Fig.~\ref{3}~\cite{yan2007breaking}.

\begin{figure}[ht]
\centering
\includegraphics[scale=0.5]{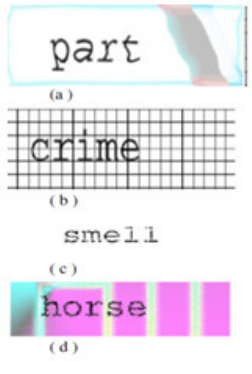}
\caption{EZ-Gimpy challenge image.}
\label{fig:figure1}
\end{figure}

\begin{figure*}[ht]
\centering
\includegraphics[scale=0.5]{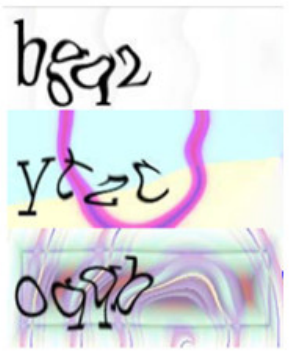}
\caption{Gimpy-r challenge image.}
\label{fig:figure2}

\end{figure*}

Gimpy-r and EZ-Gimpy are a few examples of a text-based CAPTCHA, as shown in Fig.~\ref{fig:figure1} and~\ref{fig:figure2}~\cite{moy2004distortion}. In Gimpy, an image is shown with the variegated text, and the user has to input three different texts or words by looking at them~\cite{mori2003recognizing}. reCAPTCHA is another severe form of a text-based CAPTCHA. Text-based CAPTCHA's benefit lies in its formation, where simple 4-8 letter words or arbitrarily written letters or numbers may be used. It is not very complicated for a user to recognize; however, it might not be the case as development has been done to make them robust. The deformation and noise are enhanced to make it attack-resistant, making it difficult for users to understand and recognize the letters. People with visual restraints may also find it cumbersome and fail. Moreover, if dictionary-based words are used, a huge database is required. They are also vulnerable to various attacks~\cite{yan2016simple},~\cite{ye2018yet},~\cite{uzun2018rtcaptcha}. Most of the attacks on text-based CAPTCHAs are based on OCR techniques. Therefore, the designing of a robust text CAPTCHA is inevitable~\cite{george2017generative}. Similarly, Shao et al.~\cite{shao2021robust} proposed an attack-resistant text based-CAPTCHA using adversarial examples.

\subsubsection{3D CAPTCHAs}
Many websites have employed the 2D text CAPTCHAs to protect themselves from automated bots. These CAPTCHAs are designed using random characters, symbols, and numbers in a 2D space format, as shown in Fig.~\ref{1}~\cite{woo2019design}. The most significant disadvantage with these approaches is that they can easily be cracked through automated offline post-processing attacks~\cite{woo2019design}. These approaches work by segmenting the whole CAPTCHA and then recognize each character individually. Many attempts have been made to improve security by including dots and lines across the CAPTCHAs. Deep Neural Networks (DNNs) for cracking the 2D CAPTCHAs have posed severe security threats with an accuracy of 83\%~\cite{zhou2017east}. The use of 2D text CAPTCHAs can easily be compromised with improved OCR technology. \cite{kim2019dotcha} is a new type of text-based CAPTCHA known as DotCHA, which depends on human interaction. In this approach, the user has to rotate the 3D text model to identify the correct letters. In nature, the 3D text model is the twisted and sequential 3D text having a centre pivot axis, which depicts different letters upon rotation. This type of CAPTCHA is protected from segmentation attacks. It is composed of different spheres of each letter, i.e., it is a scatter-type CAPTCHA. This type of CAPTCHA is also protected from machine Learning (ML) attacks due to its unique working way.
\begin{figure}[h]
\centering
\includegraphics[width=3in]{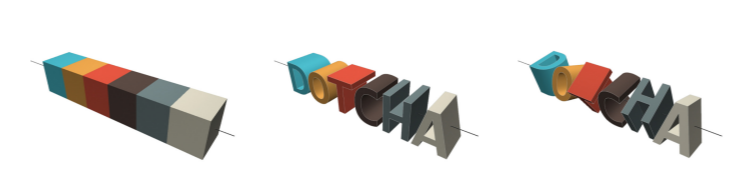}  
\caption{A DotCHA image.} \label{dot}
\end{figure}

\subsubsection{ Hand-written CAPTCHAs}
\begin{figure}[h]
\centering
\includegraphics[width=2.5in]{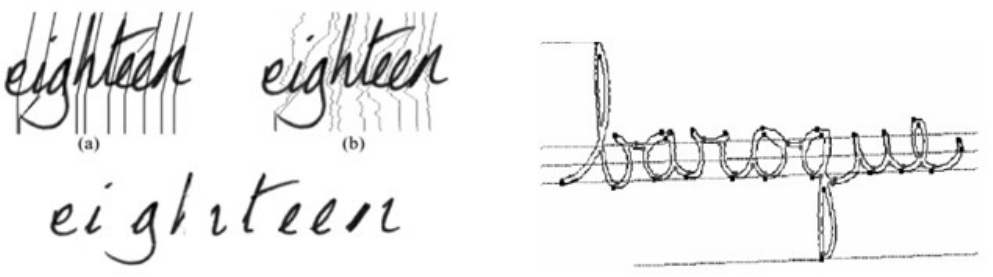}  
\caption{Showing synthesized handwritten CAPTCHA.}\label{8}
\end{figure}

Lin and Wan~\cite{lin2007style} gave an idea of a synthesized form of hand-written CAPTCHAs. GUI interface to accumulate and build samples. Mori et al.~\cite{mori2000generating} used a method that produces samples of naturally written characters. Parvez et al.~\cite{parvez2020segmentation} proposed an Arabic hand-written CAPTCHA scheme based on the ’segmentation-validation’ generation. Recently, another example is presented in~\city{kumar2022design} using Hindi language. One such example is shown in Fig.~\ref{8}~\cite{arica2001overview}.


One main advantage of hand-written CAPTCHA is that it is challenging for a computer to apply techniques like segmentation, making it grueling to recognize. However, collecting samples of hand-written words is a real problem. The size of the collection and maintaining the dataset is a large problem. It is also vulnerable to attacks, such as brute force dictionary attacks. Another issue with hand-written CAPTCHAs is the cursive writing that can also make it a little problematic for humans to identify~\cite{thomas2009synthetic}.


\subsection{Hollow CAPTCHA}
\begin{figure}[h]
\centering
\includegraphics[width=1in]{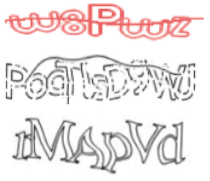}  
\caption{Showing hollow CAPTCHAs.}\label{z}
\end{figure}

Another newly emerged text-based CAPTCHA is hollow. These CAPTCHAs are deployed by many well-known websites like Yahoo!, Tencent, Sina, China Mobile, and Baidu. It uses contour lines as its` main feature in fabricating the connected hollow characters. The main aim of these contour lines is to improve the security and usability of the scheme. The standard segmentation and recognition techniques fail to break connected characters. Hollow CAPTCHA models come in different designs with various design features, such as use interference arcs, varied lengths, and thickness of hollow portions and randomly-generated contours. However, too much teeming and overlapping may confuse character pairs, compromising its usability~\cite{gao2013robustness}\cite{chen2017survey}, as illustrated in Fig.~\ref{z}.

\subsection{Non-OCR CAPTCHAs}
This section presents Non-OCR CAPTCHA schemes.

\subsubsection{Image-based CAPTCHAs}
\begin{figure}[h]
\centering
\includegraphics[width=2.5in]{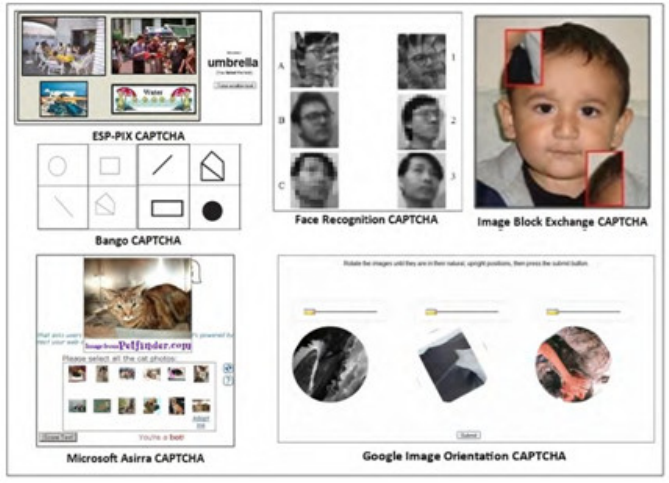} 
\caption{Different image-based challenges.}\label{4}
\end{figure}

Progress made in the making of CAPTCHAs has turned towards image-based CAPTCHAs~\cite{egele2010captcha}. The user has to identify and select a specific picture from several listed images, as shown in Fig.~\ref{4}~\cite{banday2009image}. It uses advancements in the field of image identification and blew out the inherent issues of text-based CAPTCHAs. One such example is presented in~\cite{krishna2022spiral} using adversarial perturbation and they used Convolutional Neural Network (CNN) for its security analysis.


Initially, Chew and Tygar~\cite{chew2004image} used labeled allied images in Google Image Search. After that, Ahen and Dabbish~\cite{von2004labeling} established labels in "ESP Games." However, these required classes of objects and a substantial database. Similarly, images of Kittens were utilized by Oli Warner~\cite{warner2006kittenauth}. Once again, the issue of a small database size arose. However, playing with pictures is simpler and more efficient as they are less distorted and deformed; hence they are agreeable.


\begin{figure}[h]
\centering
\includegraphics[width=2in]{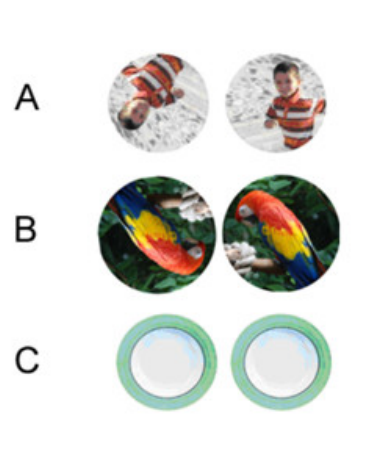} 
\caption{Images with various orientation properties.}
\label{b}
\end{figure}

\begin{figure}[h]
\centering
\includegraphics[width=2in]{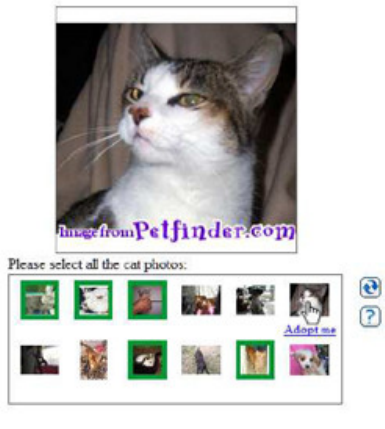} 
\caption{An ASIRRA challenge.}
\label{bbb}
\end{figure}
\begin{figure}[H]
\centering
\includegraphics[width=2in]{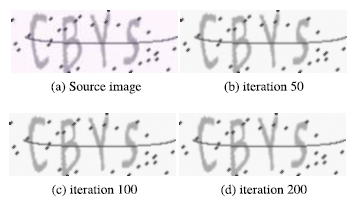} 
\caption{An example of text CAPTCHA.}\label{fig:36.png}
\end{figure}

After that, ASIRRA "Animal Species Image Recognition for Restricted Access"~\cite{elson7asirra} arrived with a
three million images database prepared by Petfinder.com. This database had two distinctive picture types (cats and dogs), which were real and innocuous, as shown in Fig.~\ref{bbb}. The test was simple as the user simply differentiates a cat from the list of dogs. This test provided 99.6\% usability relief, which the user can solve in less than 30 seconds. Compared to other CAPTCHAs, ASIRRA used attractive images; however, an algorithm and a centralized web server must be generalized and confirmed. Pictures require more storage as the database has to be maintained and displayed on-screen. There are many image-based CAPTCHAs found in literature~\cite{aleksandrovich2012image}~\cite{zhu2013image}~\cite{goswami2014facedcaptcha}~\cite{truong2020generating}. An example is shown in Fig.~\ref{fig:36.png}. Another approach is presented by Kwon et al.in~\cite{kwon2019captcha}. The same authors proposed another image-based CAPTCHA~\cite{kwon2020robust} based on adversarial example processes, such as FGSM, I-FGSM, or DeepFool. Another type of two-way image-based CAPTCHA is proposed in~\cite{bhat2020two}. The proposed scheme used a homography transformation to match two given images.


\subsubsection{ Audio CAPTCHAs}
 
 Data can be perceived in the sense of visual portrayal as well as in the form of audio~\cite{xiaojuan2010semantic}. However, image-based CAPTCHAs are less preferred over audio CAPTCHAs as some visually impaired people also utilize the Internet. Consequently, sound CAPTCHAs were created to help visually impaired users~\cite{bursztein2009decaptcha},~\cite{choi2018poster},~\cite{chithra2021scanning}. Diverse speakers speak letters or numbers after a certain time interval. The user inputs the correct number or letter in a file. Automated speech recognition (ASR) software is utilized for this functionality. Background noise is used to mystify it~\cite{bursztein2009decaptcha}. Audio CAPTCHAs may be made more robust using audio-based watermarking techniques~\cite{huang2021attacking}


  Shahreza et al.~\cite{shirali2007captcha} proposed a non-OCR CAPTCHA for visually impaired users based on Text-To-Speech (TTS) system to generate a mathematical question, and then converting it to speech. One advantage of this particular method is that automated software needs to make itself familiar with speech-based questions. In order to solve the problem and answer the questions, it needs to be smart enough. One other advantage is that a visually impaired person would not have to input long words. Just a number as a reply would be enough. Hence, it saves time as it is easy to knob.
  
   
  Haichang et al. ~\cite{haichang2010audio} used waveform diagrams for the spoken phrases or words to help the user. The phrase or word is arbitrarily made or chosen from different types of books. Such sentences appear on the screen, and the user is asked to speak them out loud. At that point, it is decided that whether the speaker is a bot or a human. This strategy analyzes sound characteristics and exploits the discrepancy between a simulated machine speech and humans. It offers a high sanctuary; however, this particular scheme's problem is settling on synthetic sound software. The issue with audio CAPTCHAs is background noise, which might affect the user's hearing acuity. We may also say that accent problems may also occur and impact recognition for non-native users. Another issue is that most audio strategies utilize audio files, which determines its span and challenge~\cite{haichang2010audio}\cite{lazar2012soundsright}. 
   
    
\begin{figure}[h]
\centering
\includegraphics[width=2.5in]{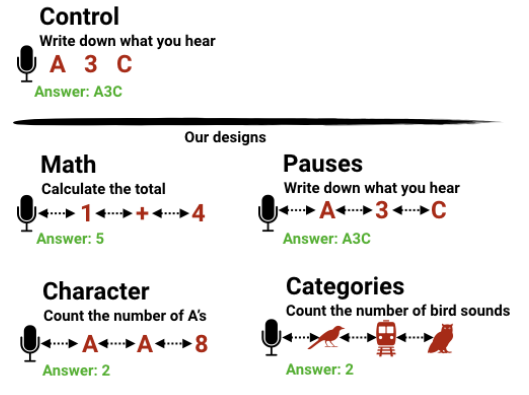} 
    \caption{Example of audio CAPTCHA.}\label{fig:37.png}
    \end{figure}
    
   Similarly, four different types of audio CAPTCHAs are proposed by Fanelle et al.~\cite{fanelle2020blind} with improved precision and speed. An example image is shown in Fig.~\ref{fig:37.png}. They conducted worldwide experimentation in three-sessions with 67 PVIs. Shekhar et al.~\cite{shekhar2019exploring} proposed a robust CAPTCHA scheme, resistant to Deep Learning (DL) and ML attacks.

\subsubsection{Collage CAPTCHAs}

\begin{figure}[h]
\centering
\includegraphics[width=2in]{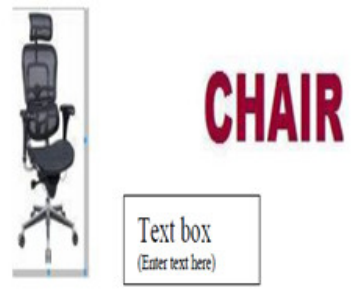} 
\caption{A collage challenge.}\label{5}
\end{figure}

CAPTCHAs mainly established on Non-OCR methods are the collage CAPTCHAs. It usually has contorted shapes that asks for the user to select a certain one. A certain Collage CAPTCHA was presented by Ritury soni~\cite{soni2010improved}. On the left of the screen, the image of objects is displayed. However, on the right side, it displays other images with the name of the left image, as shown in Fig.~\ref{5}. The user selects both the left image and the most appropriate word image from the right of the screen. The user in the given box would type the name in.Similarly, Shanker et al.~\cite{shankar2013hybrid} introduced another hybrid Collage CAPTCHA to reduce CAPTCHA attacks. 


\begin{figure}[h]
\centering
\includegraphics[width=2.5in]{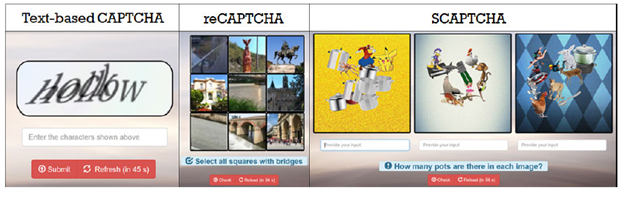} 
\caption{An example of SCAPTCHA along with a text-based and reCAPTCHA example.}\label{fig:38.png}
\end{figure}

If text and image-based CAPTCHAs are distorted, they usually displease the user and are open to attacks. The database of various pictures of a Collage CAPTCHA (animals, flags, or humans.) is displayed in different places randomly without overlapping in different places and is used after minor alternation. This objective approach is to spot the desired picture and then guess is complicated for a computer. Identification is not an easy process. The computer must be well-versed with the picture's knowledge and shapes. It has to track down the precise position of the picture displayed on the screen. Thus, resistance brews against their attacks. Such a method can be carried out on other devices with touch screens. Though it is easy to use, it still requires a database of a fixed amount of pictures that inevitably will be repeated. Another novel collage-based CAPTCHA, SCAPTCHA, is proposed in~\cite{nguyen2020secure}. An example image is shown in Fig.~\ref{fig:38.png}. It is based on different artifact segments and associated metadata, such as sizes and positions. However, users need to answer the questions associated with the CAPTCHA.

   

\subsubsection{Mouse-intrusion CAPTCHAs}
\begin{figure}[h]
\centering
\includegraphics[width=2in]{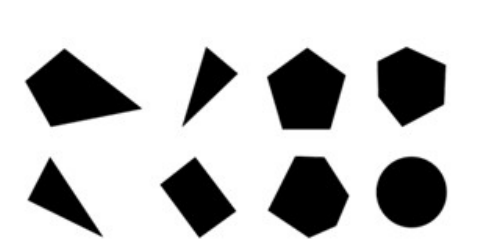} 
\caption{Mouse-intervention CAPTCHA.}
\label{mcp1}
\end{figure}

\begin{figure}[h]
\centering
\includegraphics[width=2in]{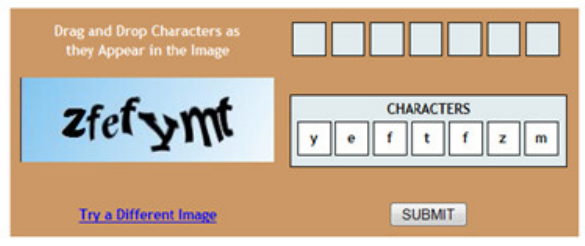} 
\caption{DND CAPTHA image.}
\label{dd1}
\end{figure}

\begin{figure}[h]
\centering
\includegraphics[width=2in]{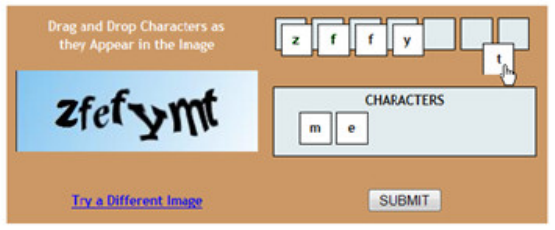}
\caption{DND CAPTCHA right or wrong placed.}\label{dd2}
\end{figure}

Among CAPTCHA breaking schemes, a text-based CAPTCHA breaking scheme is the most profitable. It is the success rate that is greater than the others. Therefore, the idea of 'Drag and Drop' CAPTCHA was introduced by Desai and Patadia~\cite{desai2009drag}. Example images are in Fig.~\ref{dd1} and Fig.~\ref{dd2}. For such kinds of CAPTCHA, mouse intrusion is mandatory. It follows the Artificial Intelligence (AI) technique. A simple mouse-intervention CAPTCHA image is shown in Fig.~\ref{mcp1}~\cite{Giangmouse}. Complicated or intricate types of CAPTCHAs usually use redirection and laundry attacks where the CAPTCHA is diverted to 'human sweatshop' after which hired humans are used so the computer can solve it~\cite{baird2005highly}. 

In~\cite{ridzuan2019image}, another click-based CAPTCHA scheme is brought forth. The mouse clicks help identify a human by analyzing the clicking on certain image components~\cite{athanasopoulos2006enhanced}. It also does that through the process of dragging and dropping a letter on the destination. The need for a keyboard in this is nil~\cite{athanasopoulos2006enhanced}. According to this approach, the users drag and drop the letters at appropriate place. In this approach, technological scrutiny is not mandatory that ends in relief for the users~\cite{chu2018bot}. 'DND' CAPTCHAs do not require any noise and chaotic backgrounds, making it apt in bandwidth use~\cite{desai2009drag}. Developed in JavaScript, HTML, and Java applets make it comprehensible, easy, and simple to design and needless bandwidth. A similar approach is proposed by Shah et al.~\cite{shah2021design}, recently.

Another click-based graphical CAPTCHA is presented in~\cite{sharma2018highly}. Such CAPTCHAs are vital and robust~\cite{desai2009drag},~\cite{chu2018bot},~\cite{cotta2019game}. However, it requires the user to properly fall through with the order and the character's exact destination. Another technique involves displaying a panel of changing colors. The user, as soon as the green color is shown, has to click the button~\cite{Giangmouse}. The mouse is also used in Bongo sentences where the clicks help the user find and then type it in the respective place~\cite{desai2009drag}. 
Likewise, game-like CAPTCHA is proposed in~\cite{kirkbridegame} for intrusion detection providing account authentication. Based on flash-gaming, another gaming CAPTCHA is proposed in~\cite{aldwairi2020efficient}.


\subsubsection{Question-based CAPTCHAs}
\begin{figure}[h]
\centering
\includegraphics[width=3.5in]{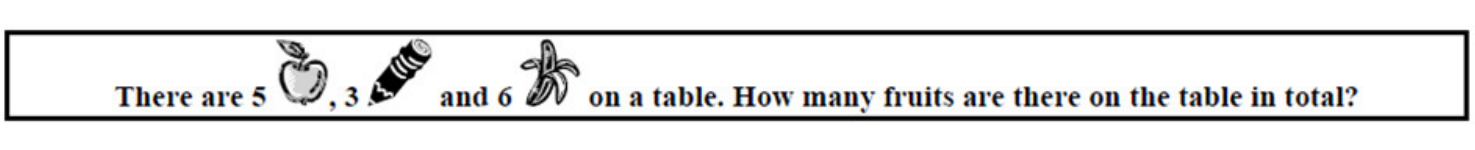}
\caption{A sample Question.}\label{qc}
\end{figure}

 A new approach uses arithmetic problems by keeping four essential abilities that humans possess~\cite{kaur2016non}. This approach involves understanding the question's text, disclosing question images, comprehending the problem, and then solving it. The arithmetic problem, to the user, is presented as an image. For example, see~\ref{qc}~\cite{shirali2007question}. Question-based CAPTCHAs are formulated to overcome the limitations of both OCR- and Non-OCR-Based CAPTCHAs~\cite{shirali2007question},~\cite{levesque2018haptic}. Gimpy method~\cite{blumcaptcha} was introduced in OCR-based CAPTCHA to take any word from the dictionary to use as CAPTCHA. Though, it was not scalable with only 850 words in the dictionary.


To lower the quality of a word to be used as CAPTCHA, a pessimal print method~\cite{coates2001pessimal} was used. However, they are also liable to hacking. The method proposed that in~\cite{shirali2006persian} would surmount the problem. This method also makes it easy, so that, non-English speakers can benefit from it. After this new approach~\cite{coates2001pessimal}, Implicit CAPTCHA~\cite{shirali2006persian} would not be in use. It is also a very costly method. To make it hard for computers to be acquainted with the speech, Text-to-Speech method~\cite{chan2003using} is used. However, the user has to make certain calculations, which people are not well-versed with it. This method is used both by the OCR and non-OCR techniques, for a bot has to go through more calculations. At times, the images can be extremely incoherent and can lead to incorrect entries by the users.


\subsubsection{Video CAPTCHAs}

\begin{figure}[h]
\centering
\includegraphics[width=2in]{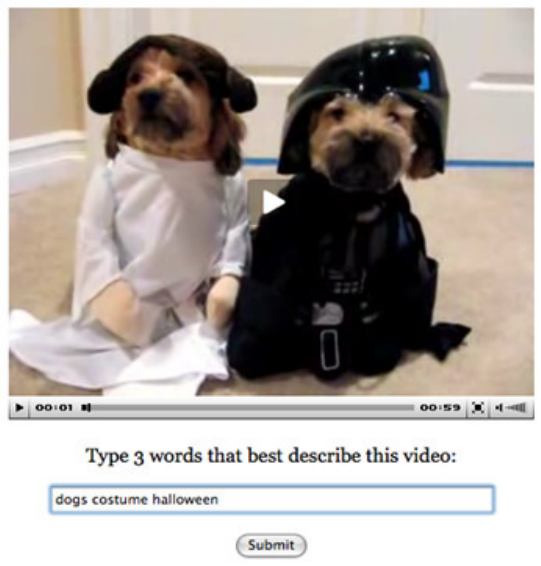}
\caption{An example of video CAPTCHA.}\label{vc}
\end{figure}

A novel content-oriented video CAPTCHA was enlisted to increase user security from 70\% to 90\% and reduce attacks with only 2\%~\cite{kluever2009balancing}. However, in this approach, video labeling is required. Every time that a video is uploaded on YouTube, it is usually tagged by it is loaded. Accordingly, every time a user plays the video, they need to type three different words, for that certain video, to describe the video. A video will be played if any of the three words match the stored database related to that video. However, the users get irritated due to wrong entries. By ensuring the Kerchkhoffs' Principle, tests are produced and checked automatically. All of the tests need to be solved with minimal exertions put by humans. Finally, it must be very challenging for the machine to be algorithmically solved~\cite{kluever2009balancing}. An example of video CAPTCHA is given in Fig.~\ref{vc}. It must be noted that this examination was conducted on a minimal scale video database; it would be ineffective to expand its size. 


Thus, for large scale video collection, the results would not be reliable. Video-based CAPTCHA solutions are discussed in~\cite{hu2013captcha}, where the text is embedded in a matter of seconds of flash video. However, the users, even without watching the whole clip, may recognize and remember characters instantly. Another problem is the equipment and software needed, such as the availability of a flash player or a compatible decoder. Therefore, the absence of such facilities can make a video playing a failure.


\subsubsection{SKETCHA: A Drawing CAPTCHA}

\begin{figure}[h]
\centering
\includegraphics[width=2.5in]{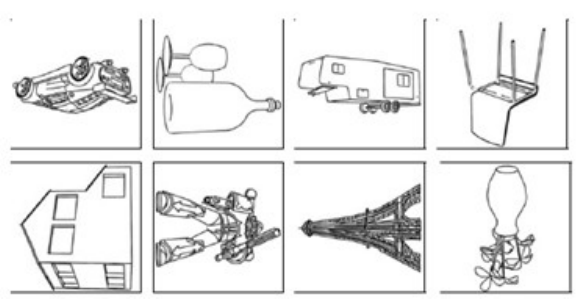}
\caption{A drawing CAPTCHA.}\label{skc}
\end{figure}

As compared to computers, humans are much better at recognizing 3D images. A new CAPTCHA called "sketcha" was devised by Ross et al.~\cite{ross2010sketcha}, with a huge database to display a 3D image in different rotations, as shown in Fig.~\ref{skc}. To bring it to 90 degrees, the user has to click. It is very accessible and friendly to the user, as a user can comprehend drawing more than photographs~\cite{gooch2004human}. However, overlapping, deformation, or addition of clutter is not required in the image, which makes a drawing CAPTCHA coherent and apparent. They are more user-friendly and have no language dependency~\cite{ross2010sketcha}. The 3D approach makes it difficult for the machine by giving different picture styles~\cite{gossweiler2009s}. Though, to bring the image to 90 degrees, several clicks may be needed, as shown in Fig.~\ref{b}. However to save the images, a database is also required.


\subsubsection{Game CAPTCHAs}

\begin{figure}[h]
\centering
\includegraphics[width=2in]{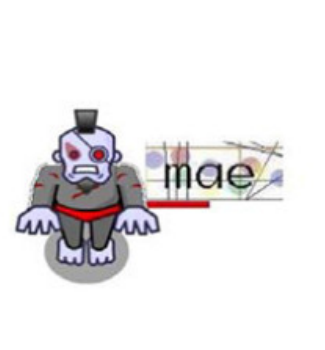}
\caption{A CAPTCHA is attached to each devil.}
\label{dev1}
\end{figure}

\begin{figure}[h]
\centering
\includegraphics[width=2in]{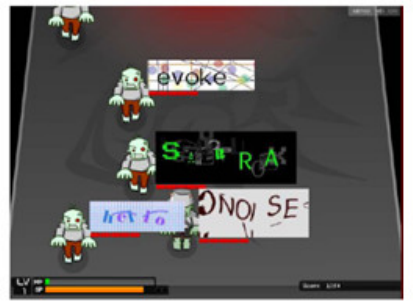}
\caption{The devil moves from upside to downside.}
\label{dev2}
\end{figure}

CAPTCHAs are mainly seen in the form of various numbers, letters, and a variety of pictures. The core reason for CAPTCHAs appearance on screen is to ward off users from bot programs. Not only do suspicious users get this unrelenting task of solving a CAPTCHA, but the genuine users also receive them. Judging this uncomfortably from the user end, Chein-Ju et al.~\cite{ho2011deviltyper} came up with a great initiative, namely "DevilType."  The maximum numbers of devils are overthrown by using CAPTCHA tests, as shown in image Fig.~\ref{dev1} and \ref{dev2}.


\begin{figure}[h]
\centering
\includegraphics[width=2in]{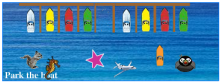}
\caption{An example of parking games.}\label{pg}
\end{figure}

\begin{figure}[h]
\centering
\includegraphics[width=2in]{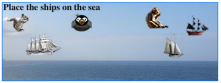}
\caption{An example of ships games.}\label{pg1}
\end{figure}

Ultimately this effective strategy helped to get exceptionally credible experiments that gave positive results. 
This technique owns two integral benefits, i.e., human's entertainment and playing become meaningful. Secondly, the game programmers take leverage when the game logo is displayed several times. 
The CAPTCHA is further segregated into different levels for users to choose from based on their convenience. In order to make CAPTCHA look a little more complicated and intrigued, more detail is added. It may be fun, however, at the same time, strenuous as well. Manar et al.~\cite{mohamed2014dynamic} proposed a relay attack-resistant Dynamic Cognitive Game (DCG) CAPTCHA. It had simple moving objects. The user plays an object matching game to pass the test, as shown in Fig.~\ref{pg} and \ref{pg1}. 
 
   
\begin{figure}[h]
\centering
\includegraphics[width=2.5in]{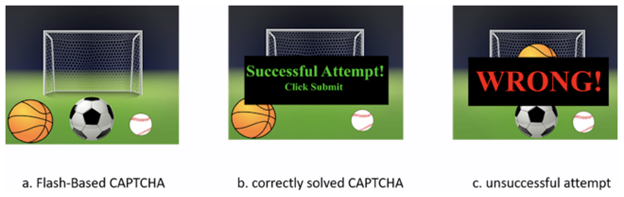}
\caption{Flash-Based Gaming CAPTCHA.}\label{fig:39.png}
\end{figure}

Aldwairi et al.~\cite{aldwairi2020efficient} proposes another CAPTCHA, as shown in Fig.~\ref{fig:39.png}. It is a flash-based CAPTCHA that is naive and easy to solve yet with the least chances of errors. It can be solved in the shortest time with easy recall capabilities. However, one needs to have technical skills to attempt the test.

\begin{figure*}[h]
\centering
\includegraphics[width=5in]{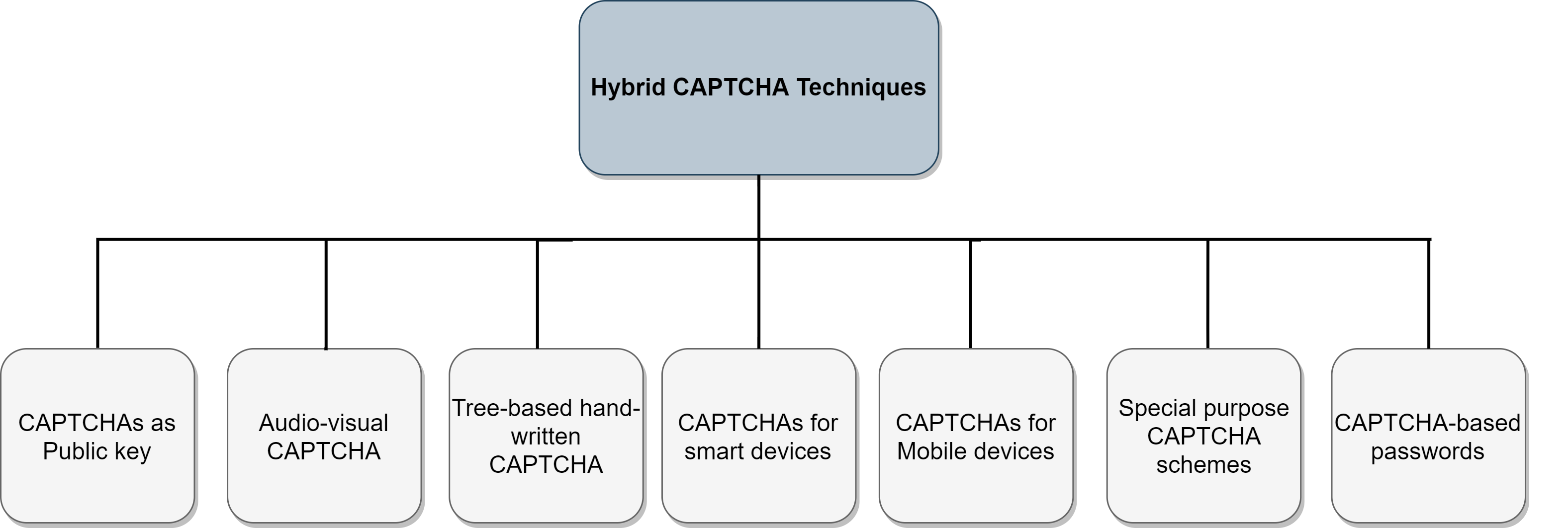}
\caption{A taxonomy of Hybrid CAPTCHA schemes.}\label{ht}
\end{figure*}

\subsubsection{Hybrid CAPTCHA schemes}
This section presents different CAPTCHA approaches based on mixed-use formations, such as text and audio, text and multimedia, and special purpose CAPTCHA schemes. Some examples of hybrid CAPTCHA may be found in~\cite{singh2020hybrid},~\cite{li2021vrcaptcha}, and~\cite{shi2021adversarial}. A taxonomy of hybrid CAPTCHA scheme is shown in Fig.~\ref{ht} and the types are discussed below:

\begin{itemize}
    \item[-] \textbf{CAPTCHAs as Public key:} All types of CAPTCHAs, whether OCR or non-OCR, are used for bot attack prevention. In Man-in-the-Middle (MitM) attack situation, they fail in securing the resources. Keeping such scenarios in mind, public-key embedded CAPTCHAs are introduced by Samir Saklikar~\cite{saklikar2008public}. To ensure the validity of such CAPTCHAs, public key installation is a mandatory process along with the surety of a secured channel. However, handshaking and public/private keys pose communication overhead in such schemes.

\item[-] \textbf{Audio-visual CAPTCHA:} 
This kind of CAPTCHA scheme is appropriate for illiterate persons. Tariq and Khan~\cite{tariq2018match} proposed a novel hybrid CAPTCHA scheme based on images and related sounds. This approach is based on the cognitive abilities of human users. It is a user-friendly approach; however, databases of images and sounds are required. Similar work examples are~\cite{shirali2007localized} and \cite{holman2007developing}.

\item[-] \textbf{Tree-based hand-written CAPTCHA:} This CAPTCHA scheme is based on the combination of text and graphics. It exploits the reading abilities of human users. A synthetic hand-written tool is used along with some random transformations to generate a tree structure test~\cite{rusu2010leveraging} randomly. However, an understanding of the tree structure is required.

\item[-] \textbf{CAPTCHAs for smart devices:} Ubiquity of Internet browsing in smart IoT devices has led to the design of CAPTCHA with special features. Due to constrained resources like memory and energy, it is inevitable to design light-weight CAPTCHAs. These schemes should promote certain characteristics of touch-screen devices~\cite{roshanbin2016enhancing},\cite{kalgi2015mobile}. Highlighting CAPTCHA~\cite{shirali2008highlighting}, multilingual highlighting CAPTCHA ~\cite{shirali2011multilingual}, CAPTCHA zoo~\cite{lin2011new} and seeSay and hearSay CAPTCHA~\cite{shirali2013seesay} are few examples of such domain.
\item[-] \textbf{CAPTCHAs for Mobile devices:} There are some specially designed CAPTCHAs for mobile devices. They make use of touch screens and the sensors embedded on the device. Such examples encompass~\cite{pritom2020combining},~\cite{acien2020becaptcha},~\cite{aburada2019implementation},~\cite{perera2019ocaptcha},~\cite{acien2020captcha}, and~\cite{acien2021becaptcha}.
\item[-] \textbf{Special purpose CAPTCHA schemes:} Special purpose CAPTCHA schemes are meant for special people, such as visually or deaf and hard of hearing people. The examples of special-purpose CAPTCHAs include~\cite{holman2007developing},~\cite{lazar2012soundsright}, and \cite{shirali2008new}. Another CAPTCHA scheme specifically presented for children in~\cite{shirali2008captcha}.
\item[-] \textbf{CAPTCHA-based passwords:} Alajmi et al.~\cite{alajmi2020password} presented a CAPTCHA-based password authentication mechanism using AI problem. Another video-based CAPTCHA is proposed in~\cite{kute2018video}. This scheme is proposed as a graphical password scheme. Muhammad et al.~\cite{mohammed20183c} also proposed a graphical password scheme using click-based CAPTCHAs. Khan et al.~\cite{khan2018new} also proposed a graphical password scheme for security using CAPTCHA. A detailed study on CAPTCHAs as graphical passwords is presented in~\cite{patel2021survey}.
\end{itemize}

 \begin{table*}
\caption{Performance analysis of different CAPTCHA Types.}
\label{tablel}
\begin{center}
\tiny
\begin{tabular}
{|p{1cm}|p{2cm}|p{1.8cm}|p{1.7cm}|p{1cm}|p{1.4cm}|p{2.3cm}|}
\hline
 \textbf{Types} & \textbf{CAPTCHA} & \textbf{Dictionary-based} & \textbf{Database-based}& \textbf{Usability} & \textbf{Vulnerability}& \textbf{Speed and Storage cost} \\ \hline
{OCR} & Text-Based & \checkmark & \checkmark & Low & High & Low \\ \hline
{OCR}& 3D-text & \checkmark & \checkmark & Medium & Medium & low \\ \hline
{OCR}&Hand-written & \checkmark & \checkmark & Low & High & Low \\ \hline 
{NON-OCR} & Image-based &  $\times$& \checkmark & High & Low & Medium \\ \hline

{NON-OCR}&Audio &  $\times$& \checkmark & Low & Low & High  \\ \hline
{NON-OCR}&Collage  &  $\times$& \checkmark & Medium & Medium & High \\ \hline
{NON-OCR}&Mouse-intervention & \checkmark & \checkmark & high & Medium & Medium  \\ \hline

{NON-OCR}&Video &  $\times$& \checkmark & High & Low & Medium   \\ \hline
{NON-OCR}&Sketcha &  $\times$& \checkmark   & High & Medium & Medium  \\ \hline
{NON-OCR}&Public key embedded &  $\times$& \checkmark  & High & Medium & Low \\ \hline
{NON-OCR}&Game &  $\times$& \checkmark 
& Low & Low & High \\ \hline
{NON-OCR}& Question-based &  $\times$& \checkmark & Low & Medium & Low 
\\ \hline
\end{tabular}
\end{center}
\end{table*}

\section{Challenges and open issues in CAPTCHAs designs} \label{coi}
A CAPTCHA is considered good if it is easy for humans to solve and resist potential attacks. However, designing such CAPTCHAs is a difficult task. A CAPTCHA is considered usable if it is easy for human users. It is considered strong if it resists computers and with a success rate as low as 0.01\%~\cite{el2011captcha}. Most of the CAPTCHA breaking techniques are designed for text-based CAPTCHAs, as it is widely used~\cite{shi2020text}. To improve the user-friendliness of CAPTCHA, the presentation of its user interface needs to be up to par. There is a myriad of qualities that affect the practicality of CAPTCHAs~\cite{brodic2020captcha}. Some of which are; font size, font style, and image size. Problems usually arise when humans do not easily recognize the font style and situations that look similar. For instance, a sizeable font is the most suitable in text-based CAPTCHA as it improves human users' clarity.

Similarly, the size is a fundamental property of image-based CAPTCHA as it also improves usability. Small images are preferable because they depreciate server-processing time, increase the download process, and inhabits smaller spaces in webpages~\cite{noorjahan2019bio}. However, large images are more precise and more descriptive for human users. They also improve the security prowess of the CAPTCHA and reduce the probability of speculations~\cite{roshanbin2016adamas}. Furthermore, the importance of sizeable images is of high importance. A resolution must be made between usability and vigor.

Audio-based CAPTCHA is quite different from its other CAPTCHA counterpart, as it does not depend on presentation. Some studies have shown that audio-based CAPTCHA might be complicated for visually impaired users. Bigham and Cavender~\cite{bigham2009evaluating} debated that most current audio CAPTCHAs are exasperating for visually impaired users. The latter use screen readers to understand CAPTCHA. The rationale behind the debate is that navigation elements cause them to lose concentration and fail to hear the CAPTCHA's beginning, even as they try to listen and answer the CAPTCHA questions. The load time can be altered about the user's network constraint to improve the image's usability- and motion-based CAPTCHAs~\cite{truong2011icaptcha}~\cite{gao2014gaming}. Conversion of the animation quality and dimensions to the grayscale can improve performance.

The inference that can be drawn from this is that intellectually disabled users are faced with more difficulties, followed by visually impaired users~\cite{berton2020captchas}. The rationale behind this inference is that intellectually disabled users do not understand CAPTCHAs. For the visually impaired, most CAPTCHAs are understood through a visual canal. Although several steps have been taken to improve CAPTCHAs' accessibility for disabled individuals, complete eradication of the problems will make CAPTCHAs more accessible by computers and robots. The use of substitute methods instead of CAPTCHAs to avoid spam should be significantly considered. Because after the analysis done on the subject, substitute methods handle accessibility hurdles better~\cite{moreno2014captcha}.

To enhance CAPTCHA usability and assist human users in identifying aim objects in noisy backgrounds, zooming. Color filtering tools should be made available. Unavailability of the listed tools can frustrate human users because of background noises in the suggested CAPTCHA, which is the same as some Unicode characters in terms of shape, color, and location~\cite{roshanbin2013survey}. The importance of colors in improving CAPTCHA usability cannot be overemphasized as color eases human identification and confound OCR systems~\cite{kaladgi2019applying},~\cite{el2011captcha}. Although colors have their usefulness, one has to be careful in their application, impairing security and usability. The perplexity associated with exploiting desirable color patterns and arrangements is one of the adverse effects on human usability. Another adverse effect is the inability to counter-attack. 

There is a constant comparison between CAPTCHA and other substitute methods~\cite{moradi2015captcha}. These substitute methods, while some may be more efficacious, user friendly, and secure than CAPTCHA. CAPTCHA is still the most common method among users, such as security administrators and business owners. Nevertheless, some substitute methods like spam detection and Askimet plugs are prevalent among users~\cite{begum2015collaborative}. The rationale behind this penchant is explained below:

\begin{itemize}
  
\item    Most administrators use third-party solutions~\cite{lopresti2005leveraging}. They do not code their CAPTCHAs from the beginning on the premise of vigor, popularity, cheap execution charge, and low management overhead. Except in crucial and delicate situations, the rationale behind the use of substitute methods is not feasible.
\item    Most administrators are unaware of the new substitute methods, and there is not enough information about them~\cite{moradi2015captcha}. Nevertheless, there are third party solutions that provide substitutes to users, unlike CAPTCHA. The substitute methods are still upcoming in terms of their diversity and popularity viewpoint.
\item The fact that most of these substitute methods are not adequately organized and packaged is responsible for most users not using them. The unavailability of these features decreases their usability and makes them informal.
\item Website owners should continue using substitute methods because of CAPTCHAs' effect on accessibility, conversion rate, usability, and user satisfaction. Managers that are aware of the consequences of using CAPTCHAs and, at the same time, unaware of substitutes have removed CAPTCHAs from their websites. They now care about their users than they do attackers~\cite{moradi2015captcha}.
\end{itemize}

\begin{table*}[!ht]
\caption{Comparison among different CAPTCHA techniques}
\resizebox{\textwidth}{!}{
\begin{tabular}{|c| c| c| c| c| c| c| c| c| c|}
\hline
 {Reference} &  {Year} &   {Type} & 
{Technique} & 
\multicolumn{6}{c}{Characteristics}\tabularnewline \cline{5-10}
& & 
&  &Clutter/Noise & Deformation & Orientation& Overlapping& User-friendly& Scalability  \tabularnewline
\hline
\cite{banday2009image} & 2009 & Non-OCR&Image&\checkmark & \checkmark & \checkmark & $\times$& Medium & Low \tabularnewline \hline
\cite{tariq2018match} & 2018 & Non-OCR&Match the sound &$\times$ & $\times$& $\times$& $\times$ & High& Low  \tabularnewline \hline
\cite{yan2007breaking} & 2007 & OCR&Google, MSN&$\times$ & \checkmark & \checkmark& \checkmark& Medium& Low  \tabularnewline \hline
\cite{moy2004distortion} & 2004 & OCR&Gimpy-r and EZ-Gimpy&\checkmark & \checkmark & $\times$& \checkmark& Low& Low 
\tabularnewline \hline
\cite{woo2019design} & 2019 & OCR&3D DotCHA&$\times$ & \checkmark & \checkmark& $\times$& Low& Low  \tabularnewline \hline
\cite{arica2001overview} & 2001 & OCR&Hand written&\checkmark & \checkmark & $\times$& \checkmark& Low& Low  \tabularnewline \hline
\cite{gao2013robustness} & 2013 & OCR&Hollow text&\checkmark & \checkmark & $\times$& \checkmark& Low& Low  \tabularnewline \hline
\cite{elson7asirra} & 2007 & Non-OCR& ASIRRA&\checkmark & \checkmark & \checkmark & $\times$& High& Low  \tabularnewline \hline
\cite{aleksandrovich2012image} & 2012 & Non-OCR&Image &\checkmark & \checkmark & \checkmark& \checkmark& High& Low  \tabularnewline \hline
\cite{zhu2013image} & 2013 & Non-OCR&Object recognition &\checkmark & $\times$ & $\times$& \checkmark& High& Low  \tabularnewline \hline
\cite{kwon2020robust} & 2020 & Non-OCR&FGSM, I-FGSM &\checkmark & \checkmark & $\times$ & $\times$& High& Low  \tabularnewline \hline
\cite{bursztein2009decaptcha} & 2009 & Non-OCR&Automated Speech Recognition (ASR) &\checkmark & $\times$ & $\times$&  $\times$& Medium & Low \tabularnewline \hline
\cite{shirali2007captcha} & 2007 & Non-OCR&Text to Speech (TTS) &\checkmark & $\times$ & $\times$& $\times$& High& Medium 
\tabularnewline \hline
\cite{haichang2010audio} & 2010 & Non-OCR&Waveform diagrams &\checkmark & $\times$ & $\times$& $\times$& High& Medium 
\tabularnewline \hline
\cite{lazar2012soundsright} & 2012 & Non-OCR&Audio &\checkmark & \checkmark & \checkmark& \checkmark& High & Medium 
\tabularnewline \hline
\cite{shekhar2019exploring} & 2019 & Non-OCR&Contemporary architecture &\checkmark & \checkmark& \checkmark& \checkmark & High& Medium   \tabularnewline \hline
\cite{shankar2013hybrid} & 2013 & Non-OCR&Hybrid &\checkmark & \checkmark& \checkmark& $\times$ & High& Medium   \tabularnewline \hline
\cite{nguyen2020secure} & 2020& Non-OCR & Object Segmentation &\checkmark & \checkmark&\checkmark& \checkmark & High & Low \tabularnewline \hline
\cite{desai2009drag} & 2009& Non-OCR & DND draggable character &$\times$ & \checkmark& $\times$&\checkmark & High& Medium \tabularnewline \hline
\cite{shirali2007question} & 2007& Non-OCR & Question-based &$\times$ & \checkmark& $\times$&$\times$ & High& low   \tabularnewline \hline
\cite{shirali2006persian} & 2006& Non-OCR &  Persian and Arabic writing &\checkmark & $\times$&$\times$ & \checkmark& Medium & Low \tabularnewline \hline
\cite{kluever2009balancing} & 2009& Non-OCR & Kerckhoffs’  Principle &\checkmark & \checkmark&\checkmark & \checkmark& Medium & Low \tabularnewline \hline
\cite{hu2013captcha} & 2013& Non-OCR &Video-based &\checkmark &$\times$&\checkmark & \checkmark& Medium& Medium  
\tabularnewline \hline
\cite{ross2010sketcha} & 2010& Non-OCR &Drawing CAPTCHA  &$\times$ &$\times$&\checkmark & $\times$& High& Medium   \tabularnewline \hline
\cite{ho2011deviltyper} & 2011& Non-OCR &A DevilType, Video Captcha  &$\times$ &$\times$ & $\times$ &\checkmark& High& Low  \tabularnewline \hline
\cite{mohamed2014dynamic} & 2014& Non-OCR & Dynamic Cognitive Game (DCG) &$\times$ &$\times$ &\checkmark & $\times$ & High & Low 
\tabularnewline \hline
\cite{aldwairi2020efficient} & 2020& Non-OCR &Flash-based &$\times$ &$\times$ &\checkmark & $\times$ & Medium& Medium  \tabularnewline \hline
\cite{saklikar2008public} & 2008& Non-OCR &Hybrid &\checkmark &$\times$ &$\times$ & $\times$ & High & Medium \tabularnewline \hline

\cite{von2008recaptcha} & 2008 & OCR&reCAPTCHA&\checkmark & \checkmark &\checkmark& \checkmark& Low& Low  
\tabularnewline \hline

\cite{gao2010audio} & 2010 & Non-OCR&Audio &\checkmark & $\times$ & $\times$& $\times$& High& Medium  \tabularnewline \hline
\cite{giang2006mouse} & 2006& Non-OCR & Mouse Intervention &$\times$ & $\times$&\checkmark& $\times$ & High& Low   \tabularnewline \hline

\cite{baird2003pessimalprint} & 2003& Non-OCR & Turing test using pessimal print &\checkmark & \checkmark& $\times$&$\times$ & High & Low 
\tabularnewline
\hline
\end{tabular}}
\label{table:comparison1}
\end{table*}

\subsection{Performance analysis}
Since time immemorial, the text-based CAPTCHA has been the most preferred type of CAPTCHA. Although it is unsafe in recent times, text-based security can still be improved using various means due to its wide usage and complexity. These means may include; use of font styles or noise. Recently, the most widely used CAPTCHA is image-based. It is advantageous in the sense that it is easier to use than its text-based counterpart. However, this CAPTCHA might be prone to adversarial attacks; there is still room for improvement. The most uncommon type of CAPTCHA is the audio/video-based CAPTCHA. Its atypicality is owed to the premise that it requires higher bandwidth and does not save time. Above all, they get broken with ease. Based on our study, the performance analysis of different CAPTCHA schemes is illustrated in Table~\ref{tablel}. Similarly, a comparison based on different CAPTCHA characteristics is presented in Table~\ref{table:comparison1}.

 We observed that the OCR-based CAPTCHAs are mostly dictionary- and database-based. The trade-off between usability and robustness is the main challenge in the design of a CAPTCHA. Text-based CAPTCHAs are the ones that are most frequently used by several websites and web-based applications. Text-based CAPTCHAs are mainly preferred due to low latency and storage costs. However, they are more vulnerable than the other CAPTCHA schemes, such as image, audio, or video. Besides, text-based CAPTCHAs are annoying to use due to usability issues. Different security features make it attack-resistant in CAPTCHA designing, such as 'font tricks,' diversified font styles, upper or lower case letters, deformation, alteration, distortion, twists, and sizes. Another such feature is the 'noise'~\cite{yan2007breaking}, which is also referred to as clutter. For noise and clutter, different shapes (such as circles, squares, or rectangles), lines (vertical, diagonal, or horizontal), colors, dots may be used for CAPTCHA security. Similarly, the objects or text may be overlapped to make CAPTCHA resilient against potential segmentation attacks.
   
On the contrary, non-OCR-based CAPTCHA tests do not require word dictionaries. However, a substantially larger database is needed to store images, videos, or sounds. They have different disadvantages, such as vulnerability, unfriendly, massive image libraries, server loads, downloading delays, and timely access~\cite{banday2009image}. However, they are less vulnerable than text-based CAPTCHAs. They are more usable than text-based because they require less noise, clutter, deformation, and overlapping. These features make them more enjoyable and easy to pass.

Similarly, audio-based CAPTCHAs are less supported as non-native listeners/users might face problems recognizing words/phrases due to accent problems. Moreover, non-native speakers may not have enough vocabulary. Downloading issues make video-based CAPTCHAs less attractive than the others. Besides, network delay and the file size may also strain a user's resources interested in the desired task and not in CAPTCHA solving.

\subsection{Recommendations for designers}
According to our study, the following are some of the ineluctable factors in designing a robust CAPTCHA.
\begin{enumerate}
\item  A good CAPTCHA must be resistant against segmentation attacks. 
\item  Deformation, fragmentation of text/images, lines, and clutter must be added to avoid recognition attacks.
\item Distortion, rotations, wrapping of text/images, and random scaling may also make it robust.
\item False boundaries of text/images are also helpful in designing a good CAPTCHA.
\item Apposite selection of multiple backgrounds and foreground colors also enhances security.
\item CAPTCHA must be generated at run-time with enough randomness.
\item Random length and combination of characters must be promoted.
\item Make it sturdy against pixel count attacks, such as by using random wrapping of text.
\item Clever use of words, phrases, and images is required.
\item To avoid database attacks, the use of a large-sized database is mandatory.
\item Feature deformation is helpful in image-based CAPTCHA designing to deter ML-based attacks.
\item A good CAPTCHA must be resilient against random guessing attacks.
\item We can confine the number of attempts and the time taken to solve a CAPTCHA.
\item Question-based CAPTCHA may be based on the logical thinking of humans rather than on computations only.
\item A combination of text and images is a good strategy.
\item Make more human interactions through mouse-clicking, screen tapping, drag-n-drop, or drop-down lists.
\item    Machine learning technology is used in both breaking and security enhancements in the CAPTCHA. An example of this technology is the adversarial and neutral style that may give better directives for designing CAPTCHA~\cite{kwon2020robust}.
\item    While the need to solve the different challenges might be pressing, some confidential information is also required~\cite{kalaichelvi2020image}. Confidential information, such as pass time, pass speed, and the operation tracks, are needed to differentiate humans from robots.
\item    Language comprehension should be paid more attention. As the ability to understand language is a required human skill to improve CAPTCHA in its entirety. Associative power is just another advanced human skill that can improve CAPTCHA.

\end{enumerate}
\begin{figure*}[!t]
\centering
\includegraphics[width=4in]{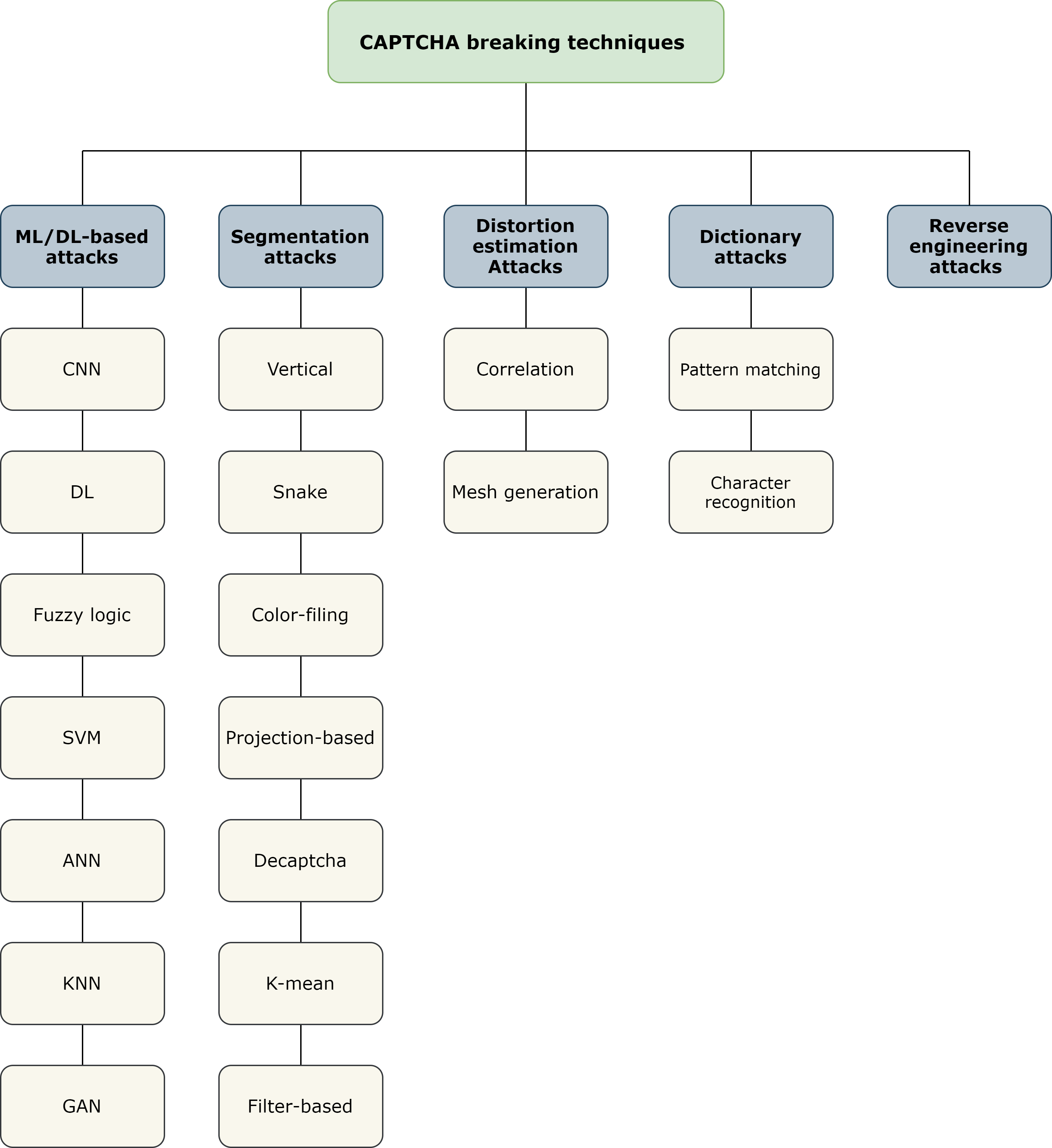}
\caption{The proposed taxonomy of CAPTCHA breaking techniques.}\label{tax2}
\end{figure*}

However, if these factors are applied eminently, the usability of the resultant challenge will decrease. As said earlier that there is a trade-off between robustness and usability. For example, adding more clutter, distortion, or overlapping will result in robust yet less readable and annoying CAPTCHA for human users. Therefore, we need to balance robustness and usability at the same time in a healthy proportion.

\section{CAPTCHA breaking techniques}\label{cb}

 Mori and Malik~\cite{mori2003recognizing} gave the idea of CAPTCHA breaking in 2003. They made recognition-based attacks on Gimpy and EZ-Gimpy CAPTCHAs. They attacked many schemes and revealed the weaknesses of different CAPTCHA schemes. This section presents different attacks made on various CAPTCHA schemes. A taxonomy of different CAPTCHA breaking techniques is shown in Fig.~\ref{tax2}.

\subsection{Segmentation Attacks}
\begin{figure}[h]
\centering
\includegraphics[width=3in]{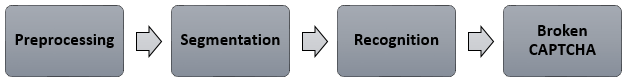}
\caption{A generic view of CAPTCHA breaking steps.}\label{st}
\end{figure}

In breaking a CAPTCHA, segmentation is the most crucial step. A CAPTCHA challenge can only be well-recognized if the segmentation step is done properly~\cite{wang2018optimized}. Chellapilla et al.~\cite{chellapilla2004using} claimed that once CAPTCHA texts are segmented well, recognizing a single character is much better in automated programs than humans. Pre-processing is done first to remove noise and clutter. Once Pre-processing has been done, the process of segmentation is started. A well-segmented image is then going through the recognition process to break a CAPTCHA scheme~\cite{chen2017survey}. A generic view of CAPTCHA breaking steps is given in Fig.~\ref{st} followed by some segmentation-based attacks made on different CAPTCHA schemes.

\subsubsection{Vertical segmentation}

  Many visual CAPTCHA schemes are attached by Chellapilla et al.~\cite{chellapilla2004using}. According to the estimate, the success rate shot from 4.89\% to 66.2\% in this segmentation. The authors stated that humans feasibly detected the twisted/distorted word into a complete word on account of this. Therefore, it is pertinent to mention that they can easily break a segmented text as well.

\begin{figure}[h]
\centering
\includegraphics[width=2in]{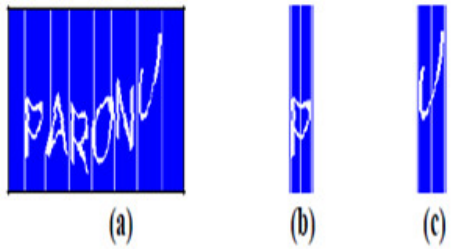}
\caption{Showing vertical segmentation.}
\label{fig:sg1}
\end{figure}
\begin{figure}[h]
\centering
\includegraphics[width=2in]{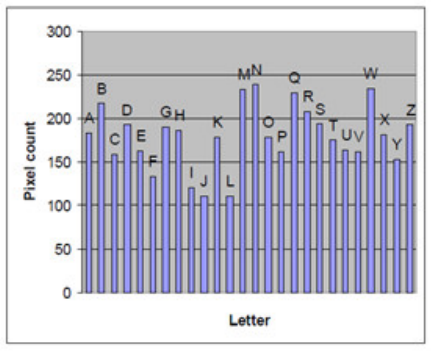}
\caption{Letter A-Z and their pixel count.}
\label{fig:sg2}
\end{figure}

Microsoft, Yahoo, and Google CAPTCHA were termed the weakest ones as humans' capability to break them were relatively higher, says Jeff Yan and Ahmed Salah~\cite{yan2007breaking}. They took six improperly sequenced letters or numerical digits from services.org CAPTCHA. Even though they were not in a sequence, overlapping words could not become possible, making humans rethink every time they click the expected digit. Hence, segmentation was possible in a vertical position. Each word was then counted in pixel table after segmentation for recognition, as seen in Fig.~\ref{fig:sg1} and~\ref{fig:sg2}.

 \begin{table}
\caption{Pixel-count lookup table.}
\label{t11}
\begin{center}
\tiny
\begin{tabular}{|l| l|}

\hline
 \textbf{Alphabet} & \textbf{Pixel-count} \\ \hline
A&183\\\hline
B&217\\\hline
C&159\\\hline
D&192\\\hline
E&163\\\hline
F&133\\\hline
G&190\\\hline
H&186\\\hline
I&121\\\hline
J&\textcolor{red}{111}\\\hline
K&\textcolor{red}{178}\\\hline
L&\textcolor{red}{111}\\\hline
M&233\\\hline
N&239\\\hline
O&\textcolor{red}{178}\\\hline
P&\textcolor{red}{162}\\\hline
Q&229\\\hline
R&208\\\hline
S&194\\\hline
T&175\\\hline
U&164\\\hline
V&\textcolor{red}{162}\\\hline
W&234\\\hline
X&181\\\hline
Y&153\\\hline
Z&193\\\hline

\hline
\end{tabular}
\end{center}
\end{table}

Table~\ref{t11} represents the lookup table used in letter-pixel count. The success rate of this approach was nearly 100\%. In contrast, the strategy fails to function in letters with the same number of pixels (highlighted red in Table~\ref{t11}), especially when they overlap. Additionally, suppose the same pixel refers to more than one character. In that case, it paves the way for ambiguity and malfunctioning of the segmentation algorithm.


\subsubsection{Snake segmentation}

\begin{figure}[h]
\centering
\includegraphics[width=3in]{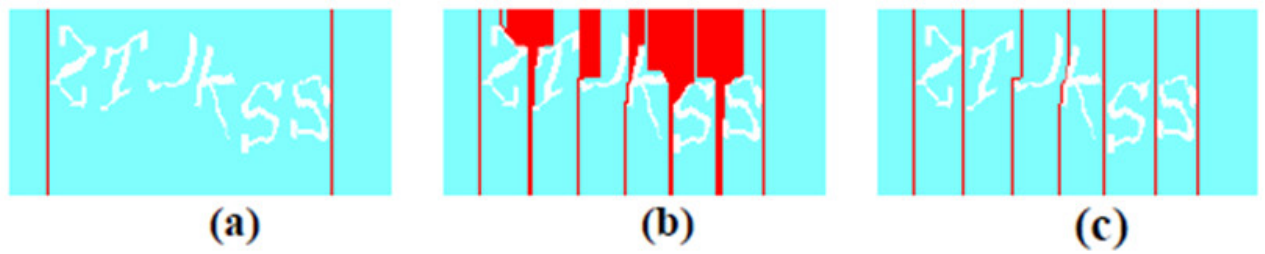}
\caption{Snake segmentation:(a) Pre-processing,(b) Prior to segmentation(c) After segmentation.}\label{snk}
\end{figure}

To get rid of the simple segmentation, snake Segmentation~\cite{yan2007breaking} was used. There is no such rule in Snake Segmentation that it has to be vertical only. As shown in Fig.~\ref{snk}, the snake can move forward and backwards and up and down to break a picture. Standardization is done to see the letter, and that happens when the segmentation is done. A success ratio of 96\% in the examples given and 99\% on pictures selected randomly was attained. The images with alphabets, words, or numbers are difficult to break down because they are either joint or overlapping images.


\subsubsection{Color filling Segmentation}

\begin{figure}[h]
\centering
\includegraphics[width=2.5in]{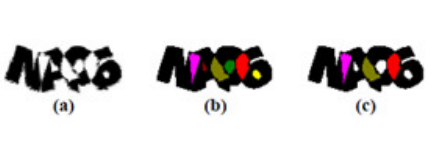}
\caption{Component identification:(a) Original CAPTCHA(b) Background removal(c) Loop removal.}
\label{fig:cl1}
\end{figure}

Color filling segmentation was used for pictures that contain only two colors, one of them being the background~\cite{el2010robustness}. The other color is used for the text; therefore, the foreground. The foreground letters are dispersed, and to form an image, different colors are combined, and each pixel's count in the segment is done. The count done here is then compared with the count that is in the pixel table. If the count falls short, it is added with its adjoining right and left segment to figure out the correct letter or digit, as shown in Fig.~\ref{fig:cl1}.

\begin{figure}[h]
\centering
\includegraphics[width=1.5in]{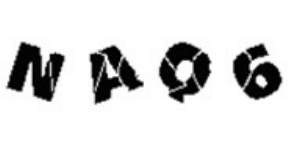}
\caption{Final recognized characters.}
\label{fig:cl2}
\end{figure}

If there are many arcs placed, then this method is not successful. Megaupload~\cite{el2010robustness} made a new segmentation resistance that was not like Google, Yahoo, or even Microsoft. There are a total of 4 digits and letters that are placed horizontally. Some of the parts have overlapped in them. They are eliminated and are only in two colors, black and white. For color segmentation to be done, the black and white parts are taken out one after another. They are added to recognize the white components, and any clutter or background loops are cut off. They are added together to make a letter in the end. In 100 sample tests, the success ratio is 82\% with sufficient attack speed; 400 different samples had a success ratio of 78.25\%. However, serious problems can be caused if two or more of the characters are joined to the black part, as shown in Fig.~\ref{fig:cl2}. 

If the loop is not recognized accurately, a problem can occur. Another similar attack is launched on image-based CAPTCHAs, IMAGINATION, in~\cite{zhu2010attacks}. They used a color-edge and line-segmentation detection mechanism. One hundred and nine (109) images of the said CAPTCHA scheme were brought together online, and with a success ratio of 74.31\%, they broke 81 CAPTCHA images.



\subsubsection{Projection-based segmentation}

\begin{figure}[h]
\centering
\includegraphics[width=3in]{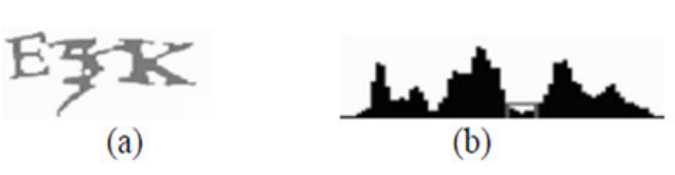}
\caption{(a) Test image (b) Projection-based image.}\label{pr}
\end{figure}

To break the CAPTCHA scheme, there are two basic methods; identification and segmentation~\cite{huang2008projection}. The procedure of identification acknowledges a character when it has been segmented well. Letters and clutter were identified separately through label segmentation, and Chellapilla~\cite{chellapilla2005computers} also used it. The segmentation method fails of the difference is not much, and the characters tend to break to eliminate noise. As shown in Fig.~\ref{pr}, projection-based segmentation was used by Huang et al.~\cite{huang2008projection} for Yahoo and MSN CAPTCHAs. The technique used by them had a rate of success of 9\% more than the Chellapilla. As it helps to get rid of the clutter and noise, that is the advantage it has over Chellapilla. To analyze the CAPTCHAs, this technique helps to give a new dimension.


\subsubsection{DeCAPTCHA}

Bursztien and Bethard~\cite{bursztein2009decaptcha} attacked eBay audio CAPTCHA. They represented a proposal, 'DeCAPTCHA', in order to break the sound-based CAPTCHA schemes. 'CAPTCHA chorus' is made by the scraper to make such an attack. To break Sound CAPTCHAs, we can use Sphinx~\cite{walker2004sphinx}, which is a state-of-the-art sound recognizer. A graph is made when the audio is changed to features to decode the words and letters that are said. For this purpose, auditory and language modules are needed. The DeCAPTCHA broke a total of 75\% of the sound-based CAPTCHAs, and the downloads were limited for this reason.


\subsubsection{K-Means layered segmentation}

\begin{figure}[h]
\centering
\includegraphics[width=3in]{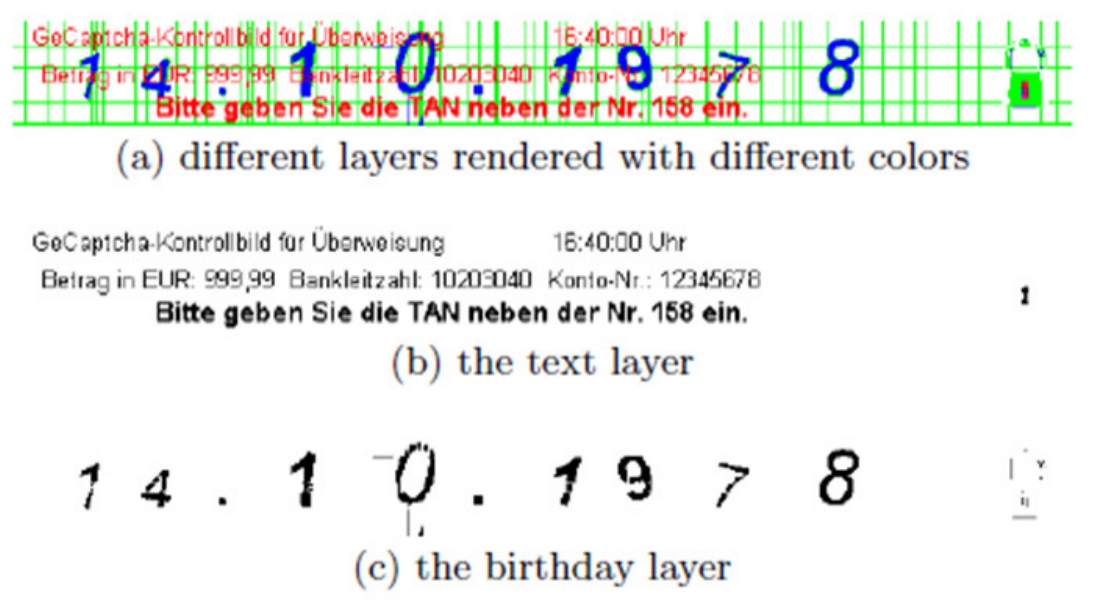}
\caption{Segmentation results of a Ge-Captcha image.}\label{eb}
\end{figure}

As shown in Fig.~\ref{eb}, there are transactions conducted in the e-banking on the CAPTCHA verification bases to avoid attacks by the bot. They help prevent unauthorized access and keep it safe from attacks by a Man-in-the-Middle (MitM). The pattern detected is processed by a tool, namely K- mean layered segmentation. There are three layers of e-banking CAPTCHAs, the first one having grids that are placed erratically. User birthday entrenched is the second layer, with each letter and digits having font style, location, and random alternation. The final uppermost layer has a transaction date on it. 
There are several steps to carry the attack out. To better the e-banking CAPTCHA, there are good lines provided. The rate of success is around 100\% for breaking e-banking CAPTCHA. Li and Khayam~\cite{li2010breaking} suggested not to use the e-banking CAPTCHA and use hardware security accomplishment instead.



\subsubsection{Filter-based segmentation}

\begin{figure}[h]
\centering
\includegraphics[width=3in]{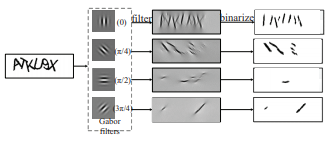}
\caption{character component extraction using Gabor filters.}\label{fl}
\end{figure}

A low cost yet powerful attack based on Log-Gabor filters is proposed in~\cite{yan2016simple} to break various designed text-based CAPTCHAs. The attack shows a 5\% to 77\% success rate with less than 5 seconds on average to break a  single instance of text CAPTCHA. They also compared Log-Gabor filters accuracy with other filter-based segmentation schemes, such as 2D Gabor filters and steerable filters. The results show that the Log-Gabor filter segmentation outperformed on character extraction than the two, as shown in Fig.~\ref{fl}.  

\subsection{ML- and DL-based attacks on CAPTCHAs}
Automated processes can also break CAPTCHAs with ML and speech recognition algorithms. Golle~\cite{golle2008machine} showed that ML-based attacks have a 10.3\% success rate on Asirra. Similarly, in~\cite{zhu2010attacks}, an ML-based attack is made on image-based CAPTCHAs with a 40\%  success rate. Saroha et al.~\cite{saroha2021strengthening} used a backpropagation-based mechanism for CAPTCHA recognition. The following are some ML-based techniques used in CAPTCHA breaking.

\subsubsection{Convolutional Neural Network (CNN)}
 CNN is used for analyzing visual imagery, and they are good in areas like image classification~\cite{krizhevsky2012imagenet}. The other variants of CNN, such as Regions with CNN features (RCNN) and Fast-RCNN, Faster RCNN, YOLO, and SSD, are integrated with improved security features and object detection accuracy~\cite{weng2019towards},~\cite{du2017captcha},~\cite{liu2017cnn}, and~\cite{sachdev2020breaking}. CNN is equipped with a better mechanism with enhanced accuracy and time complexity. However, CNN's establishment requires much effort, including pre-trained staff members and much homework~\cite{shu2019end}. Furthermore, it requires a large quantity of high-quality data along with a dependable neural network. Hence, it requires a high initial cost~\cite{wang2018optimized}. A CNN-based Chinese text-based CAPTCHA breaking mechanism is proposed in~\cite{jia2018approach}. Similarly, Stephen and Lu~\cite{dankwa2021efficient} assessed cyber security vulnerabilities by breaking CAPTCHAs using CNN. A semi-supervised DL-based (CNN) approach is used in~\cite{bostik2021semi} for breaking common CAPTCHA schemes.
 
\begin{figure}[h]
\centering
\includegraphics[width=3.5in]{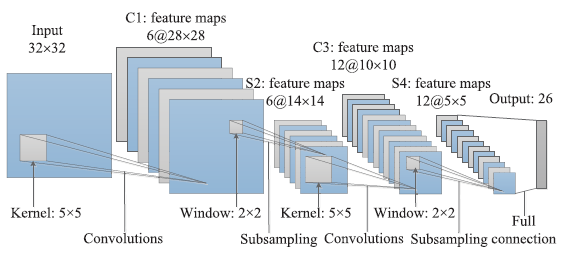}
\caption{CNN network structure.}\label{clw}
\end{figure}

 \cite{chen2018attack} proposed non-redundant merging with an accurate filling scheme for cracking the hollow CAPTCHA. In solid characters, they applied the nearest-minimum neighbor merging algorithm for gaining the correct individual characters. They used CNN to obtain the final recognition results with a success rate of 97.50\%  for individual characters and 91.00\% for single images. A depiction is shown in Fig.~\ref{clw} \cite{weng2019towards} discussed three different proof-of-concept attacks for selection-based CAPTCHAs, slide-based CAPTCHAs, and click-based CAPTCHAs. They break the CAPTCHA of 10 real-world security systems. They also discussed this method's working efficiency with previous methods, i.e., image recognition service and humans involved in the CAPTCHA solving process. The success rate of the scheme is 85.72\%-99.65\%.

 A Deep Convolutional Neural Network (DCNN) was presented to break such CAPTCHAs based on a confusion class. However, it has low recognition accuracy. Therefore, the selective-learning confusion class was introduced to overcome this problem for text-based CAPTCHAs in~\cite{chen2019selective}. The process involves a two-stage DCNN framework based on the confusion relation matrix. The partitioning algorithm improved 1.4\%-39.4\% for confusing characters for confusion class DCNN and the validation and training algorithm. Another DCNN-based attack is made on image-based CAPTCHAs in~\cite{singh2018machine}. They took 25,000 test images, and the accuracy rate is as high as 95.2\%. Similarly, Zhang et al.~\cite{zhang2021development} propounded Captchanet as a classification mechanism for Chinese character-based CAPTCHAs with better accuracy than other DL-based mechanisms. Another CAPTCHA recognition scheme is presented by Wand and Shi~\cite{wang2021captcha} based on CNN with  99\%, 97.84\%, and 98.5\%, recognition rate.

\subsubsection{Deep learning}

Deep learning is considered the most dominating method among different existing techniques from the last few years. Due to this method, the existing techniques have developed significantly better recognition than humans. Deep neural networks usually are integrated with all the levels of features and end-to-end based classifiers involving the multiple layer fashion enriched with the numbers of depths~\cite{vatanen2013pushing}. They have the degradation issue when these start to converge, which increases with the increasing depth and results in reduced accuracy~\cite{he2016deep}. This technology is employed in different recognition fields, such as image, text, and audio recognition. These techniques, such as CNN, RNN, LSTM-RNN, are some to mention, which are used in CAPTCHA recognition~\cite{chen2017survey}.

As discussed earlier, the CNN approach recognises characters' images without extracting the features and is strong in displace, deformation, and scaling. The research~\cite{goodfellow2013multi} shows that the recognition of CAPTCHA with a distributed deep convolutional neural network has an accuracy of 99.8\%. However, it cannot be combined with context information recognition due to time constraints. Hence, feedback and time parameters based RNN  was proposed for handling time series data. However, despite that, this scheme has a dispersing gradient issue. Therefore, the LSTM approach was proposed~\cite{liang2011highly}, using the 2D LSTM-RNN for recognizing CCT CAPTCHAs with the accuracy of 55.2\%. This approach was successful in obtaining the relevant information on the horizontal as well as the vertical axis. This paper proposes the generic, simple, and fast attack approach for decoding text CAPTCHA with a superposition challenge. Similarly, Tang et al.~\cite{tang2018research} proposed a generic, simple, yet fast approach for decoding CAPTCHA schemes with the superposition. The deep learning-based breaking of roman-text-based character CAPTCHA with the success rate of 10.1\%-90.0\%. The scheme also attacked three Chinese CAPTCHAs with success rates of 93.0\%,
32.2\% and 28.6\%, respectively.
\begin{figure}[h]
\centering
\includegraphics[width=3.5in]{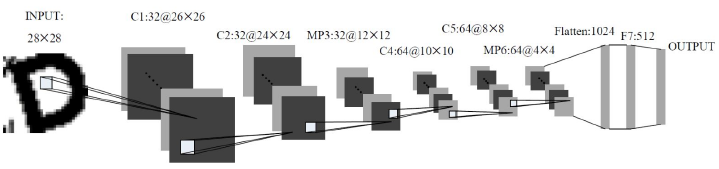}
\caption{LeNet-5 feed-forward architecture.}\label{dp}
\end{figure}
Sivakorn et al.~\cite{sivakorn2016robot} designed a novel low-cost attack that controls the semantic annotation of images through deep learning technology on image-reCAPTCHA and Facebook image CAPTCHA schemes. The accuracy rate of the said scheme is 70.78\% for image reCAPTCHA and 83.50\% for Facebook image CAPTCHA.

\subsubsection{Support Vector Machine (SVM)}
SVM is also considered a proficient classification technique. It is based upon Statistical Learning Theory with many advantages, such as robustness, accuracy, and effectiveness, even with few training samples. They are non-parametric binary classifiers by nature. It works by separating classes through a hyper-plane. The kernel function uses a key, which decodes the original features and converts them into high-dimensional space in a nonlinear way, enhancing the separability for data~\cite{chen2017survey}. They are widely applied in machine vision fields like character, hand-writing digitization, text recognition, and even satellite image classification~\cite{anthony2007classification}.

\begin{figure}[h]
\centering
\includegraphics[width=3in]{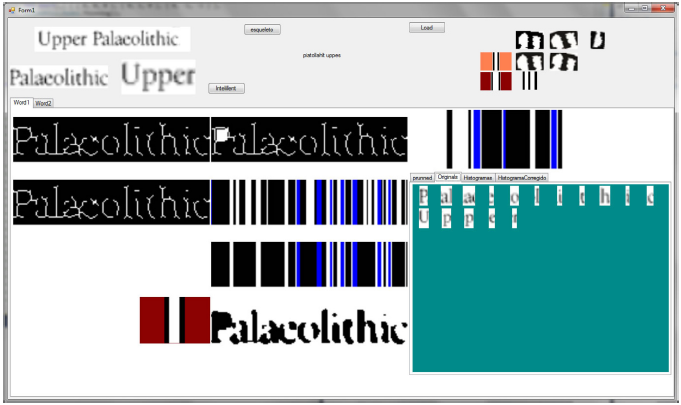}
\caption{The GUI visualization of character segmentation and SVM-based recognition used in~\cite{starostenko2015breaking}.}\label{sv}
\end{figure}

Starostenko et al.~\cite{starostenko2015breaking} presented an innovative approach for solving the automatic segmentation and recognizing reCAPTCHAs and other text-based CAPTCHAs using an SVM-based classifier. Different font sizes having different waving patterns and rotation in this approach collapse in random over-lapping characters. The suggested segmentation process is founded on three-color bar character encoding, which separates the letter in reCAPTCHA. The average segmentation rate was up to 82\%, while the SVM classifier success rate was almost 94\%. However, this technique fails in decoding the reCAPTCHAs of 2012. Another such technique is proposed in~\cite{golle2008machine}, broke image-based CAPTCHAs using SVM with an 82.7\% accuracy rate.

\subsubsection{Fuzzy logic}
Out of other techniques in clustering, the most commonly used algorithms is the Fuzzy C-Means (FCM). Fuzzy logic is widely used for image segmentation. The mechanism is a well-known clustering approach that uses numerical data (k-means method) or large datasets, i.e., a tiered method for solving dissimilarity matrices \cite{gao2012divide}. It works by partitioning the finite collection into sets of \textit{c} fuzzy clusters. It results in the formation of clusters involving every element to a certain degree. These elements can also belong to the different clusters at the same time, i.e., soft clustering. A fuzzy logic-based mechanism is propounded in~\cite{fonseca2000using} by Fonseca et al. for recognizing geometrical shapes interactively.

 Their algorithm can easily recognize elementary geometrical shapes like triangles, rectangles, circles, and ellipses. It observes and collects the geometrical features of different shapes from the database and combines them to recognize the final geometrical figure. They used the perimeter and area to recognize features that separate all the geometrical figures, as given by Grahm's Algorithm~\cite{o1998computational}. After this, each fuzzy set belonging to each feature was used for the recognition of shapes. Sasi et al.~\cite{sasi1994handwritten} studied the application of fuzzy set theory for the classification of hand-written characters sets. Initially, all the hand-written characters were classified into 4x4 matrix cells. They were analyzed and stored in the database under different categories. Another edge corner-based segmentation–recognition mechanism is proposed by Nachar et al.~\cite{nachar2015breaking} using fuzzy logic for breaking different types of CAPTCHA schemes. The success rate for eBay, Wikipedia, ReCAPTCHA, and Yahoo was 68.2\%, 76.7\%, 62.5\%, and 57.3\%, respectively.   

\begin{figure}[ht]
\centering
\includegraphics[width=2.5in]{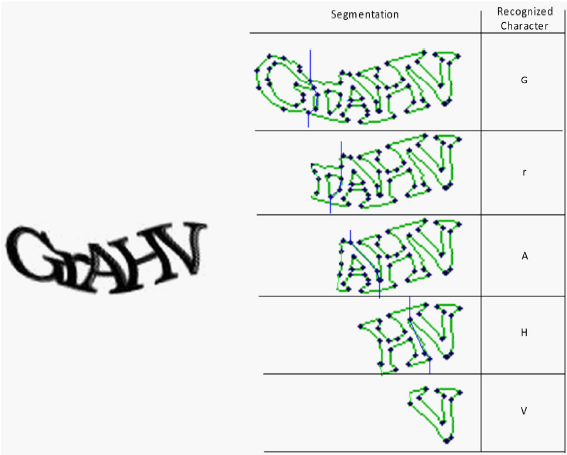}
\caption{Edge corner-based fuzzy logic segmentation and recognition.}\label{fz}
\end{figure}

\subsubsection{Artificial neural networks (ANN)}
For image recognition, different neural network approaches have been proposed. All interconnected units, called artificial neurons, are trained to solve complex problems. Each neural network consists of hundreds of layers, which are the same as the human brain. However, neural networks require intensive training with thousands of samples along. Moreover, the quality of training data and feature extraction also play an important role in final recognition~\cite{chen2017survey}.
 Simard et al.~\cite{simard2003best} came forward with their application for visual documentation analysis. Another ANN-based mechanism is proposed by Koerich et al.~\cite{koerich2005recognition} using discrete Hidden Markov Models (HMM) for unconstrained hand-written words' recognition and verification. Furthermore, Kaur et al.~\cite{kaur2012artificial} showed that an ANN system could also be used to recognize the character. Barve et al.~\cite{barve2012optical} showed that an optical character recognition system designed with ANNs could have high noise tolerance. In~\cite{liang2011highly}, a BPNN employed a cross entropy method using Recurrent Neural Network (RNN) to calculate the network's performance. The overall precision for the CCT CAPTCHAs of Taobao is 51.3\%. It is 27.1\% for MSN and 53.2\% for the eBay. However, the main problem is the extraction character feature of high quality, limiting the final results.

\subsubsection{Generative Adversarial Networks (GAN)}
A GAN uses a generative network for generating mock examples and a discriminate network for distinguishing the produced examples from the actual scenarios. For this purpose, back-propagation is used to train both networks; after some training iterations, the generators give better mock samples. In contrast, the discriminator becomes better using flagging synthetic data~\cite{ye2018yet}. GANs showed the extraordinary results in image~\cite{isola2017image}\cite{zhu2017unpaired} and natural language~\cite{yu2017seqgan}\cite{li2017adversarial} processing tasks. However, due to the technique's freshness, no literature work is available where GANs were used for solving the text-based CAPTCHAs. Therefore,a GAN-based bot is proposed in~\cite{ye2018yet} for solving text-based CAPTCHAs. The bot uses training data and can solve a wide range of text-based CAPTCHA schemes. The deviation from the past ML-based attacks requires few newly created CAPTCHAs for building the solver. It is achieved by learning the synthesizer and then generating training data for building a solver base. It them refines the solver through transfer learning on real CAPTCHAs data sets. The main advantage of this approach is that it requires very little human intervention for learning.

\begin{figure}[!t]
\centering
\includegraphics[width=3.5in]{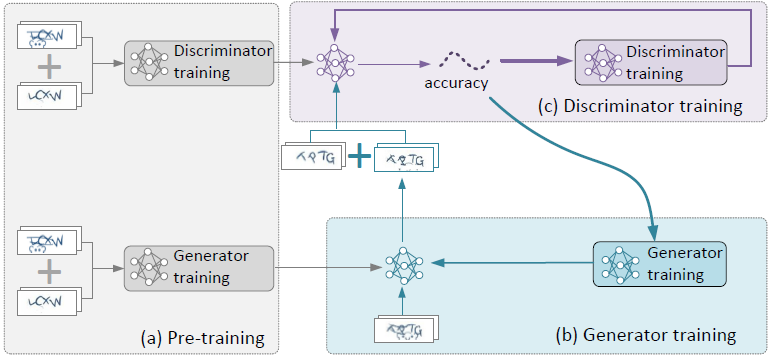}
\caption{ A GAN-based Pre-processing model  proposed in~\cite{ye2018yet}.}\label{gn}
\end{figure}

\subsubsection{K-nearest neighbours (KNN)}
KNN is designed on the category having the nearest \textit{k} samples for determining the category of a test sample. KNN is the most commonly used recognition classifier for breaking text-based CAPTCHA systems. However, the accuracy of KNN classification depends on the size of samples and the value of \textit{k}. 
In~\cite{yan2016simple}, several approaches like SVM, BPNN, template matching, CNN, and KNN were tested. The results showed that KNN achieved the highest success rates for most schemes, whereas  CNN showed better results in the shortest time.

\subsection{Distortion estimation technique}
\begin{figure}[!t]
\centering
\includegraphics[width=1.5in]{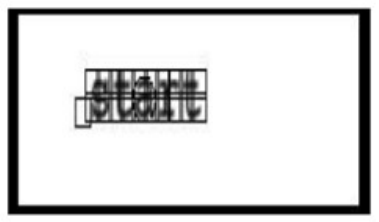}
\caption{Identifying 24 minipatches in the image.}
\label{fig:co1}
\end{figure}

\begin{figure}[!t]
\centering
\includegraphics[width=1in]{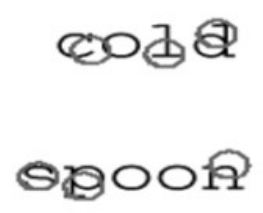}
\caption{Gimpy-r challenge with background removal.}
\label{fig:co2}
\end{figure}

Distortion-estimation techniques break EZ-Gimpy and Gimpy-r CAPTCHAs. A 99\% accuracy on Ez-Gimpy and 78\% on Gimpy-r CAPTCHAs is projected by Gabial Moy et al.~\cite{moy2004distortion}. They used a dictionary of 561-word in breaking Ez-gimpy and a four-letter word dictionary for Gimpy-r CAPTCHAs. The correlation is used in matching words by cores and minipatches identification, as shown in Fig.~\ref{fig:co1} and~\ref{fig:co2}. The correlation mechanism gave high success and accuracy rates; however,  it failed in identifying high distorted images. Also, the dictionary used in breaking was small, which made execution ineffective. However, on the increased size of slicing, the execution time increased, as well.


\subsection{Dictionary attacks}
\begin{figure}[!t]
\centering
\includegraphics[width=2.5in]{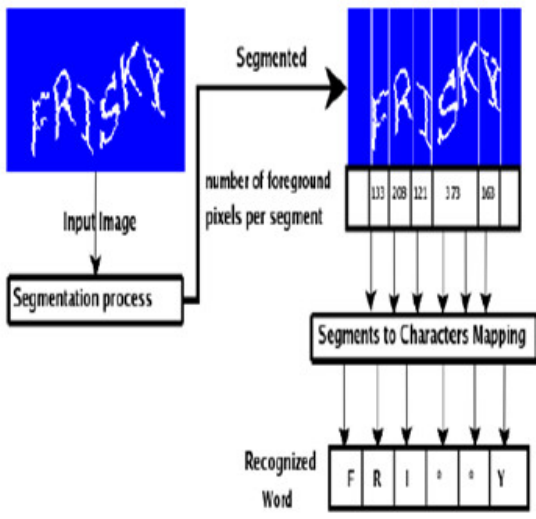}
\caption{A basic attack: a failing example where segmentation is not possible between 'k' and 's'.}
\label{d1}
\end{figure}

\begin{figure}[ht]
\centering
\includegraphics[width=2.5in]{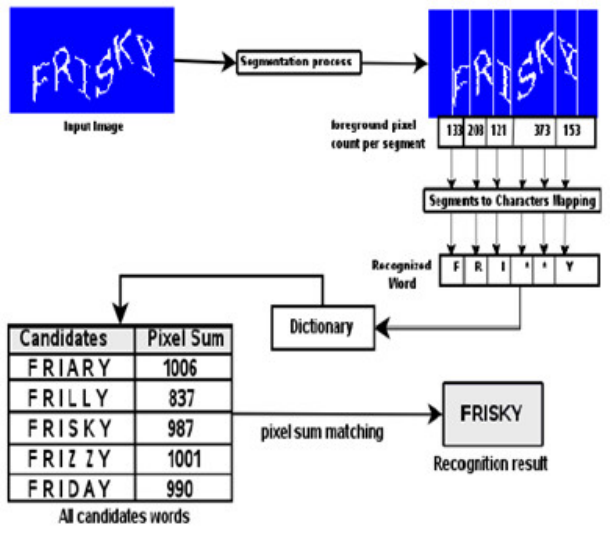}
\caption{A basic attack with dictionary enhancement.}
\label{d2}
\end{figure}

In networks, a user is likely to encounter a Denial of Services (DoS) attack~\cite{tariq2019security}. While in an allotted time restriction, if the amount of guesses is incorrectly made, the account becomes locked. The issue is encountered by legitimate users who end up facing service poverty. To get the better of these kinds of issues, EZ-Gimpy and Gimpy CAPTCHAs had been initiated. Mori and Malik~\cite{mori2003recognizing} struck the mentioned variations of CAPTCHAs, turning it uncertain by virtual brute force strikes. The procedure of this kind of attack is displayed in Fig.~\ref{d1} and~\ref{d2}. 


Text-based CAPTCHAs are where brute force attacks are formed, especially those containing a file. If not, then a database is linked alongside them. Yan and Salah~\cite{yan2007breaking} used gullible arrangement identification techniques to shatter a category of visual CAPTCHA. The highlighted issue of evident segmentation and pixel counts was conquered with the use of brute force strikes. A word is primarily sent through connection and, after that, from recognition. However, the scheme malfunctions if disconnected or linked letters are greater than 3 in numbers or all displayed or entered letters are joined. Brute force strikes are doable for words that are entered in the English language. However, unsystematic character/inclusion of digit CAPTCHA cannot be fractured by such variations of strikes.

\subsection{Reverse engineering method}

\begin{figure}[!t]
\centering
\includegraphics[width=3.5in]{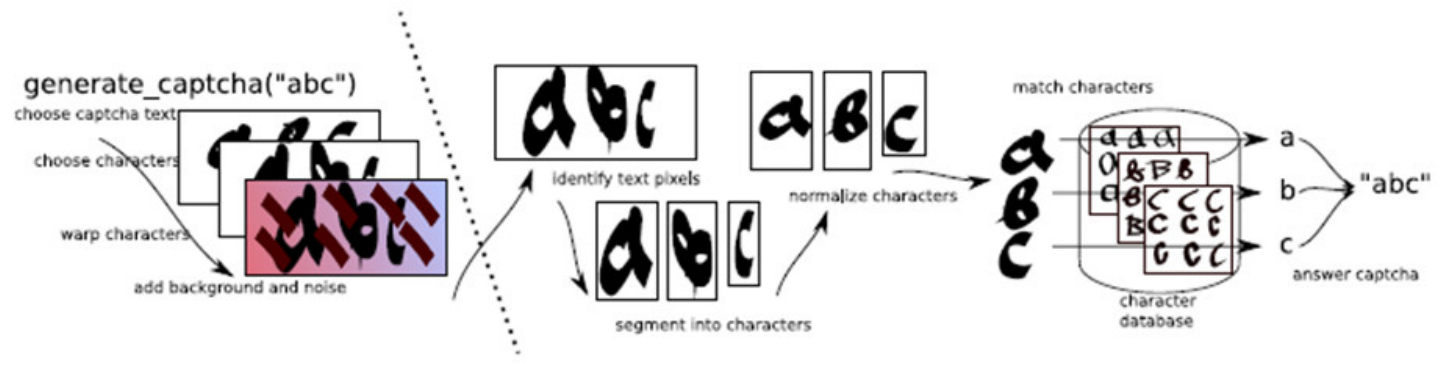}
\caption{Model of computer CAPTCHA solving.}\label{re}
\end{figure}

Hindle and Michael.~\cite{hindle2008reverse} came up with the suggestion of reverse engineering for breaking a particular CAPTCHA scheme, as shown in Fig.~\ref{re}. They compared CAPTCHAs' weakness after breaking them to highlight the way to make them more robust. However, it uses a bit map illustration solely for CAPTCHAs based on text.

The accuracy level of the technique used for PHPBB CAPTCHA is 99\%. For Rogers CAPTCHAs, it is 95\%, for Piratebay is 61\% and for Watercap CAPTCHA is 27\%. This perspective has provided adequate proposals for building CAPTCHAs more applaudable. However, it is unable to take in the use of sound and visual-based CAPTCHAs.


\section{Challenges and open issues in CAPTCHA breaking techniques} \label{coi1}
As mentioned before, there are different types of CAPTCHAs, such as text, image, audio, or video. They provide security against malevolent bot attacks. Unfortunately, they have the same fate and prone to breaking attacks~\cite{adams2015detecting}. Their security and usability, at the same time, is a challenging task. The current study on CAPTCHA reveals that although much work has been done on their security, however, there are still design vulnerabilities that make them insecure~\cite{Zhang2019survey},~\cite{tang2018research},~\cite{Chow2019}. Previously (in this paper), it is observed that several attacks are made on well-known CAPTCHA schemes, such as breaking of reCAPTCHA used extensively by Google. That is because there is no standardized mechanisms and practice for designing robust CAPTCHAs. Therefore, by exploiting their vulnerabilities, the hackers manage to break CAPTCHAs readily to harm the target source.

Another important reason behind breaking CAPTCHAs is to find their design vulnerabilities to bring security improvements in their design. There are different steps that are involved in CAPTCHA breaking, such as pre-processing~\cite{sun2019captcha}, image processing~\cite{wei2019research}, pattern matching~\cite{sonwalkar2020captcha}, ML~\cite{alqahtani2020image}, are just to name a few. The pattern matching-based attacking technique was widely used in the beginning. The downside of this attack is that it requires manual instructions to extract features for different CAPTCHA types. On the contrary, segmentation attacks are the most typical type of attack. Conversely, this method is not suitable for real-time attacks because it requires a prolonged duration, affecting the breaking speed. There are other end-to-end attacking models, which use ML-based models, for example~\cite{zi2019end},~\cite{li2020end}, and~\cite{shu2019end}. However, they require more hardware and computational resources. For instance, a deep learning approach to recognize a CAPTCHA in its entirety requires high hardware requirements.
 \begin{table*}[!ht]
\centering
\caption{A summary of different CAPTCHA breaking techniques}
\label{vvvv}
\scriptsize
\begin{tabular}{|c|c|c|c|c|c|c|}
\hline
  {\textbf{Attack type}} &  {\textbf{Algorithm}} & \multicolumn{5}{c|}{\textbf{Summary}} \\ \cline{3-7}
    &  & \textbf{Ref.} & \textbf{Year} & \textbf{Accuracy} & \textbf{Samples} & \textbf{CAPTCHA} \\ \hline
 {Machine learning} & Distortion estimation & \cite{moy2004distortion} & 2004 & 99\% & 561 & Text\\ \cline{2-7}
   & K-nearest neighbors & \cite{yan2016simple} & 2016 & 77.2\% & \textbf{--} & Text\\ \cline{2-7}
    
   & GAN & \cite{ye2018yet} & 2018 & 87.4-90\% & 500 & Text\\ \cline{2-7}
    & CNN & \cite{chen2017survey} & 2018 & 91\% & 200 & Text\\\cline{2-7}
    & CNN & \cite{wang2018optimized} & 2018 & 99.75\% & \textbf{--} & Text\\ \cline{2-7}
    & ML & \cite{zhu2010attacks} & 2010 & 40\% & 800 & Image\\ \cline{2-7}
    & SVM & \cite{golle2008machine} & 2008 & 82.7\% & 13K & Image\\ \cline{2-7}
    & ML & \cite{singh2018machine} & 2018 & 95.2\% & 25K & Image\\ \cline{2-7}
     & DCNN & \cite{goodfellow2013multi} & 2014 & 99.8\% & 100k & Text\\ \cline{2-7}
  
      & RNN & \cite{liang2011highly} & 2011 & 27-53.2\% & \textbf{--} & Text\\ \cline{2-7}
      & DL & \cite{sivakorn2016robot} & 2016 & 70.78\% & 700 & Image\\ \cline{2-7}
      & DL & \cite{sivakorn2016robot} & 2016 & 83.50\% & 700 & Image\\ \cline{2-7}
       & Fuzzy logic & \cite{nachar2015breaking} & 2015 & 57.3-76.7\% & \textbf{--} & Text\\ \cline{2-7}
    \hline
 {Segmentation}
  & Snake & \cite{yan2007breaking} & 2007 & 94\% & 100 & Text \\ \cline{2-7}
   & Vertical & \cite{yan2007breaking} & 2007 & 99\% & 100 & Text \\ \cline{2-7}
 & Divide and Conquer & \cite{gao2013robustness} & 2013 & 83\% & 1k & Text\\ \cline{2-7}

 & Colour filling  & \cite{el2010robustness} & 2010 & 78\% & 400 & Text\\ \cline{2-7}
  & Color-edge \& line segmentation & \cite{zhu2010attacks} & 2010 & 74.31\% & 109 & Image\\ \cline{2-7}
    
  & Projection & \cite{huang2008projection} & 2008 & 75\% & \textbf{--} & Text \\ \cline{2-7}
& K-Means & \cite{li2010breaking} & 2010 & 100\% & 100 & Text\\ \cline{2-7}
      \hline
  {Dictionary}
    & OCR & \cite{yan2007breaking} & 2007 & 99\% & 100 & Text\\ \cline{2-7}
 & TM/OCR & \cite{wang2018optimized} & 2018 & 98.98\% & \textbf{--} & Text\\ \cline{2-7}
     \hline
 {Distortion Estimation}
 & Reverse engineering & \cite{hindle2008reverse} & 2008 & 93\% & 100 & Text\\ \cline{2-7}
  \hline
 {Software-based} & Decaptcha & \cite{bursztein2009decaptcha} & 2011 & 75\% & 26k & Audio\\ \cline{2-7}
        \hline
\end{tabular}
\end{table*}
Under the light of our findings, The following points must be considered in designing attacks for breaking a CAPTCHA scheme:
\begin{enumerate}
\item  The CAPTCHAs, especially text-based, are mostly broken using segmentation attacks. Therefore, segmentation-strategies must be revisited to tranche the overlapped text/objects.
\item A larger size of dictionary/database must be kept updated to launch dictionary attacks. Besides, pixel statistics should also be revised from time to time.  
  \item  The end-to-end attack is one of the most foremost directives for CAPTCHA breaking~\cite{shu2019end}. These attacks are mostly Dl-based, which pose particular challenges in terms of required resources. Thus, there needs to have a lightweight and more convenient DL-based CAPTCHA-breaking approach.
  \item  There is a problem associated with ML techniques, which requires a significant number of construed datasets. This problem cannot be disregarded, so much so that a less dependent model on the training dataset is required.
  \item  An exclusive trained system is required to break different CAPTCHA. A more non-specific network that can be moved to multiple means is needed because dataset labeling is expensive.
  \item An ML-based algorithm, such as DL, is a threat to CAPTCHA security. Hence, a new CAPTCHA defense system is needed to be equipped enough to counter it~\cite{tang2018research}.  Multi-layer feature representations and abstractions from various kinds of data are the most critical purpose of ML. With deep learning, a convolution neural network was applied to image classification. A myriad of other tasks since 1989, such as~\cite{zhao2004genetic},~\cite{zhao2019object},~\cite{zhao2016corrupted},~\cite{wang2008classification},~\cite{zhao2012completed}, and~\cite{lecun1989backpropagation} have been proposed. AlexNet, in recent times, has enhanced the architecture of CNN and notably to improve the classification effect~\cite{Zhang2019survey}. It has also become the customary trainers for CNN on GPUs~\cite{xu2020survey}. 
  \item There is a limit in ML technology when involved with AI image processing, such as; symmetry~\cite{he20162016} and adversarial examples~\cite{osadchy2016no} and~\cite{szegedy2013intriguing}. According to Osadchy et al.~\cite{osadchy2017no}, CAPTCHA designed using concerning deep learning limitations is a viable future exploration~\cite{osadchy2016no}.
  \item  Web access and software generation are the two ways in which researchers get the CAPTCHA samples. The importance of standard test databases as a requirement for breaking in database-based CAPTCHAs, such as traditional test databases, is necessary for an improved text-based CAPTCHA.
  
  \item A database is required to provide dependable testing and training data for analytical work. Therefore, to gather, categorize, arrange, and initiate a CAPTCHA database is of high importance.
  \item  There is a myriad of feature changes in a CAPTCHA. Presently, the efficacious identification of CAPTCHAs by a classifier can only be made when training sets and set types belong to the same class. Therefore, it is quite onerous and necessary to design a standard classifier to identify different CAPTCHA types.
  \item  The high success rate that has been achieved by Text-based CAPTCHA breaking is astounding. Sadly, Crowding Characters Together (CCT) strings have been widely used in the text-based CAPTCHA, causing a snag in the results~\cite{kwon2019captcha}. This drawback is due to the problem associated with the segmentation-free CAPTCHA identification; a situation that requires immediate attention. This problem may be solved using deep learning because of its new ideas and techniques.
  \item 
Machine learning has been found to deliver quality results than other conventional techniques~\cite{hernandez2023breaking}. For instance, CNN and its improved methods, Deep Belief Networks (DBN)~\cite{ben2020dbn}, RNN~\cite{li2020alec}, and Deep Reinforcement Learning (DRL)~\cite{uzkent2020efficient}, are the most applied techniques. However, they still are not frequented in CAPTCHA identification. Although careful analysis, the application of fusion and correlation between the different deep learning types is not thorough. There is still a need for more effective and efficient learning models to enable ease in CAPTCHA identification.
  \item  There has been an increase in identification results' dependability due to better breaking CAPTCHA techniques. Therefore, there is a need to certify the correct rejection, dismissal, and improve the accuracy rate of recognition. Rejection as a concept is not well-known in the CAPTCHA recognition field. It is, however, quite clear that it can be developed.
  \item  In the extraction of features automatically, using ML, characters with similar characteristics are handily muddled. Practically, the accuracy of feature extraction and training methods can be improved by various ML-based techniques, such as DL and its variants~\cite{xu2020survey}.

\end{enumerate}
\begin{figure}[!t]
\centering
\includegraphics[width=2.5in]{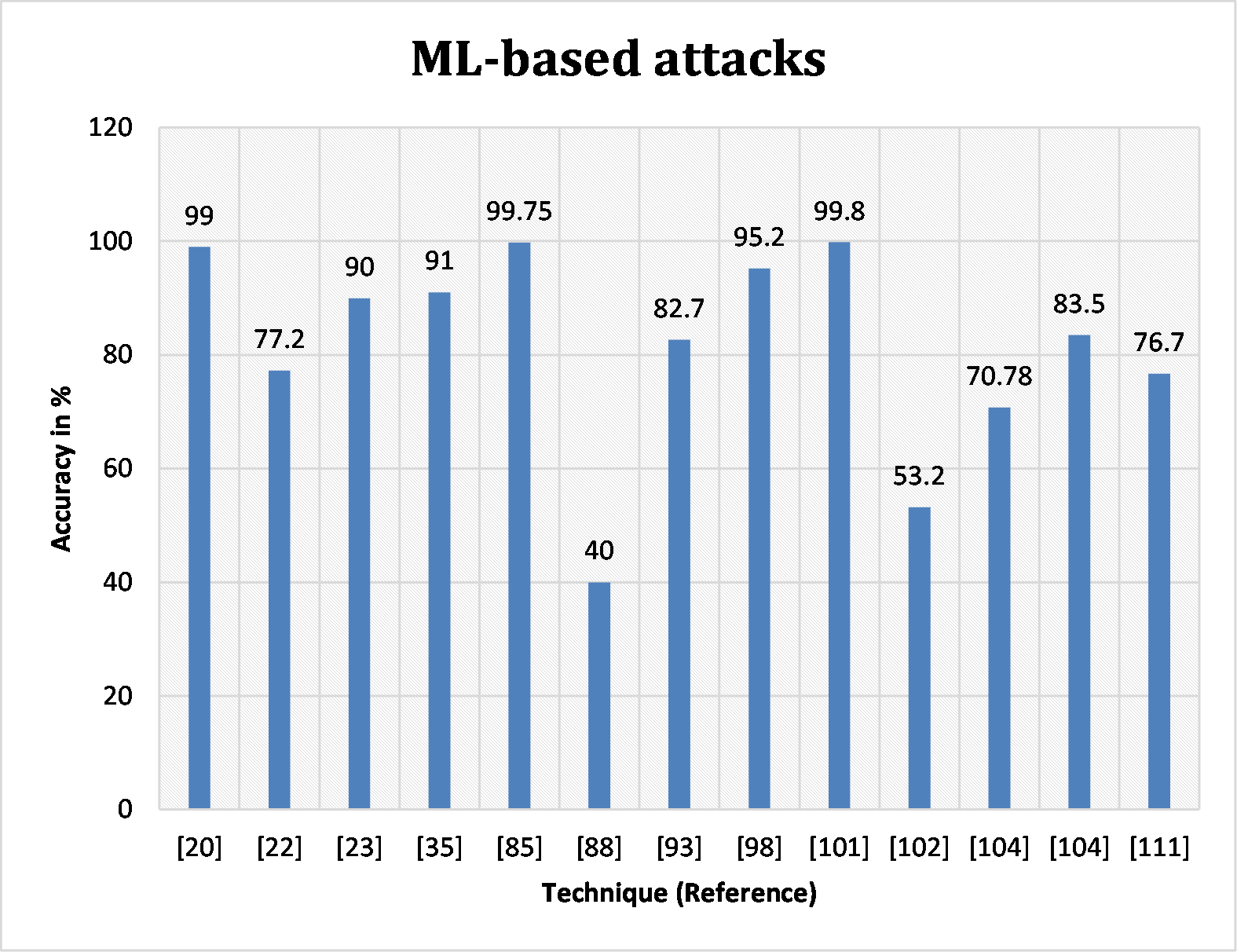}
\caption{Accuracy comparison of different ML-based attacks.}\label{g11}

\end{figure}

\begin{figure}[!t]
\centering
\includegraphics[width=2.5in]{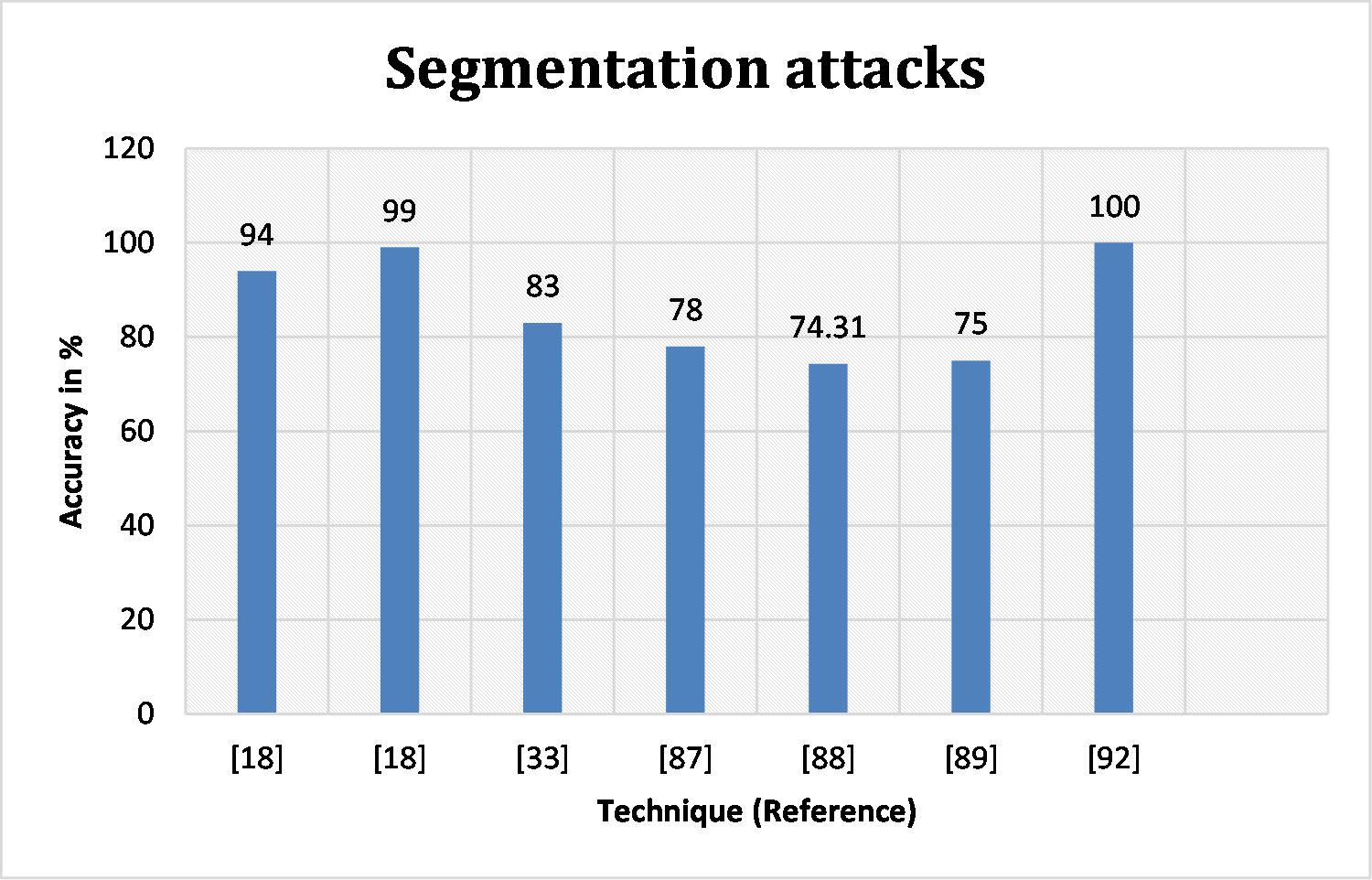}
\caption{Accuracy comparison of different segmentation attacks.}\label{g2}

\end{figure}

\begin{figure}[!t]
\centering
\includegraphics[width=2.5in]{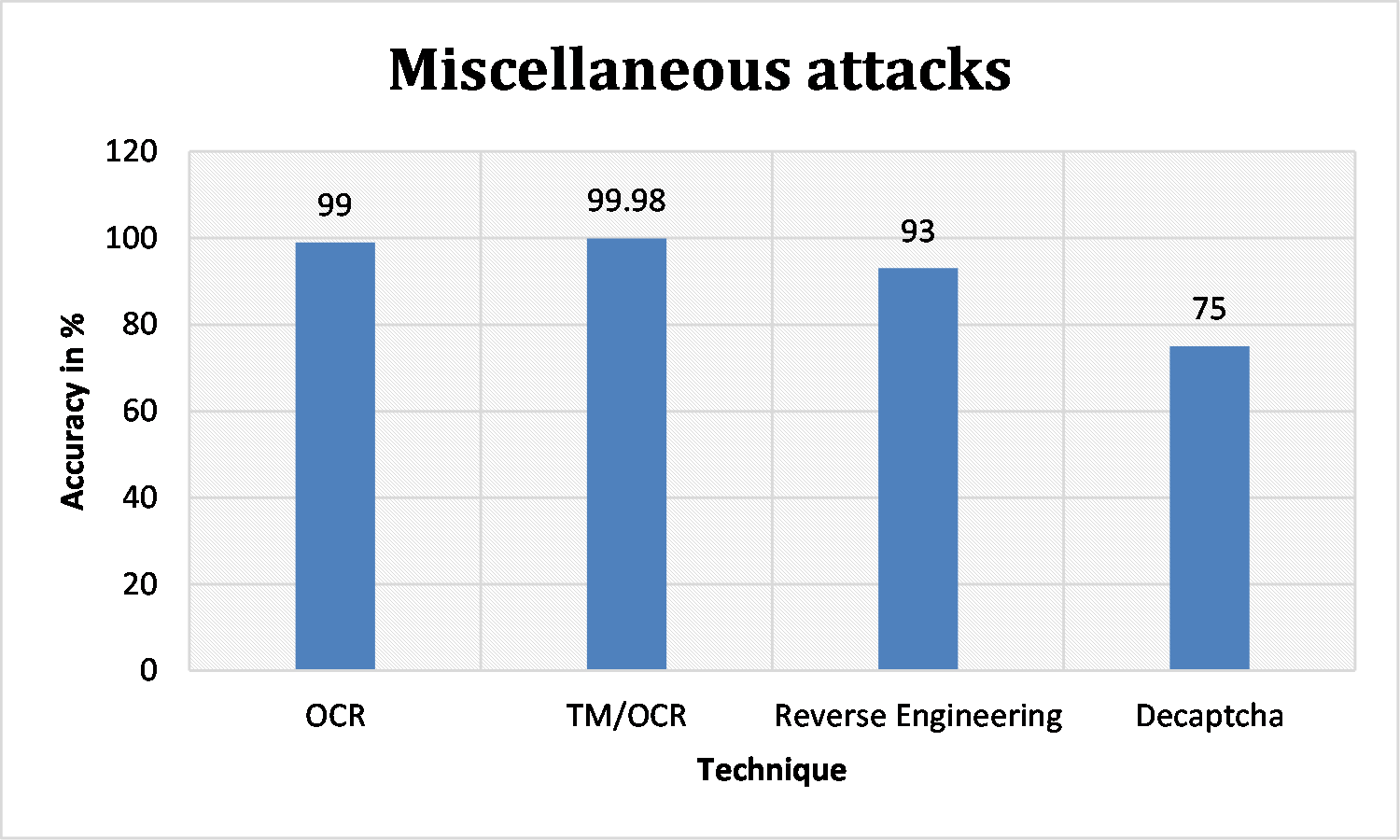}
\caption{Accuracy comparison of miscellaneous attacks.}\label{g3}
\end{figure}

\subsection{Performance analysis of CAPTCHA breaking techniques}
In keeping with our study, Table~\ref{vvvv} presents a summary of the comparison among different state-of-the-art CAPTCHA breaking mechanisms. As said earlier, ML-based attacks are widely used in CAPTCHA breaking, out of which DL variants are the most prominent. As depicted in Fig.~\ref{g11}, DCNN~\cite{goodfellow2013multi} outperformed other ML techniques with a 99.8 percent accuracy rate. Similarly, we make a comparison of segmentation-based attacks, shown in Fig.~\ref{g2}. Among all, the K-Means-based~\cite{li2010breaking} segmentation attack outperformed others with accuracy as high as 100 percent. And then comes the vertical segmentation attack~\cite{yan2007breaking} with a 99 percent accuracy rate. However, both the breaking techniques are old and were used a decade before. Thus, they might not work equally well on today's advanced CAPTCHA schemes. We also compared other breaking techniques, whose depiction is shown in Fig.~\ref{g3}.

\section{Conclusion}\label{con}
Deciding whether the user is a human or a malicious bot program is challenging for a web service provider. CAPTCHA is a security mechanism used against automated malicious bot programs. There are different types of CAPTCHA based on their formation and structure for distinguishing humans and machines apart. This survey provided an in-depth analysis of various CAPTCHA schemes categorized as OCR and non-OCR CAPTCHA schemes. We analyzed the usability and the design loopholes of each CAPTCHA type. The most crucial issue in CAPTCHA designing is the deformation and noise, compromising its usability and making it annoying for a human user. This study spotlit the design vulnerabilities and suggestions to the designers for making CAPTCHA schemes more user-friendly yet attack-resistant. Besides, it also highlighted the challenges and open issues in CAPTCHA designing.

However, CAPTCHA security is still a question. They are subject to attacks. There are different braking techniques, such as segmentation, ML-, and DL-based attacks. Among others, it is noted that ML-based, especially DL-based breaking schemes, are more dominantly practiced. This survey showed the vulnerabilities of different CAPTCHAs schemes based on the state-of-the-art CAPTCHA breaking schemes. It also provided an in-depth analysis of these breaking schemes.

Moreover, it highlighted challenges and open issues in breaking the various CAPTCHA schemes. It also provided a discussion on CAPTCHA breaking techniques that how better breaking techniques might be designed. The study would help in the usable and attack-resistant CAPTCHA designing. In the future, we aim to design a robust and user-friendly image-based CAPTCHA that is also scalable.

\bibliographystyle{unsrt}
\bibliography{sample-base}


\end{document}